%% file: DistribComp.tex
\documentclass[draftcls, letterpaper, onecolumn]{IEEEtran}
\AtBeginDvi{}
\usepackage[cmex10]{amsmath}
\allowdisplaybreaks[1]
\usepackage{amssymb,amsthm}
\usepackage{mathrsfs}
\usepackage[noadjust]{cite}
\usepackage[dvipdfmx]{graphicx}
\usepackage[dvipdfmx]{color}
\usepackage{bm}

\theoremstyle{definition}
\newtheorem{theorem}{Theorem}
\newtheorem{lemma}{Lemma}
\newtheorem{corollary}{Corollary}
\newtheorem{remark}{Remark}
\newtheorem{definition}{Definition}

\newtheorem{example}{Example}
\newtheorem{proposition}{Proposition}
\newcommand{\eqtri}{\triangleq}

\newcommand{\Ex}{\mathbb{E}}%Expectation E[X]
\newcommand{\abs}[1]{\left\lvert#1\right\rvert}%Abstract |x|
\newcommand{\plimsup}{\text{p-}\limsup}%p-limsup
\newcommand{\oH}{\overline{H}}%sup-entropy
\newcommand{\code}{\Phi}
\newcommand{\error}{\mathrm{P}_\mathrm{e}}%Error Probability
\newcommand{\svc}{q}% Slowly Varying Constant; the constant q of a slowly varying source
\newcommand{\SW}{\mathsf{SW}}
\newcommand{\id}{\mathsf{id}}
\newcommand{\fl}{\mathsf{fl}}
\newcommand{\vl}{\mathsf{vl}}
\newcommand{\EQ}{\mathsf{Equiv}}

\newcommand{\cC}{\mathcal{C}}
\newcommand{\cD}{\mathcal{D}}

\newcommand{\cR}{\mathcal{R}}
\newcommand{\cS}{\mathcal{S}}
\newcommand{\cT}{\mathcal{T}}
\newcommand{\cU}{\mathcal{U}}
\newcommand{\cV}{\mathcal{V}}

\newcommand{\cX}{\mathcal{X}}
\newcommand{\cY}{\mathcal{Y}}
\newcommand{\cZ}{\mathcal{Z}}

\newcommand{\bmZ}{\bm{Z}}

\newcommand{\bmf}{\bm{f}}

\newcommand{\bms}{\bm{s}}

\newcommand{\bX}{\bm{X}}
\newcommand{\bY}{\bm{Y}}
\newcommand{\bx}{\bm{x}}
\newcommand{\by}{\bm{y}}
\newcommand{\hx}{\hat{x}}
\newcommand{\hy}{\hat{y}}
\newcommand{\hbx}{\hat{\bm{x}}}
\newcommand{\hby}{\hat{\bm{y}}}
\newcommand{\hvarphi}{\hat{\varphi}}
\newcommand{\hpsi}{\hat{\psi}}
\newcommand{\barr}{\bar{r}}
\newcommand{\GF}{\mathsf{GF}}

\title{A Dichotomy of Functions in Distributed Coding: An Information Spectral Approach}
\author{Shigeaki Kuzuoka~\IEEEmembership{Member,~IEEE} and Shun Watanabe,~\IEEEmembership{Member,~IEEE}
\thanks{The work of S.~Kuzuoka was supported in part by JSPS KAKENHI Grant Number 26820145.}
\thanks{S.~Kuzuoka is with the Faculty of Systems Engineering, Wakayama University, 930 Sakaedani, Wakayama, 640-8510 Japan, e-mail:kuzuoka@ieee.org.}%
\thanks{S.~Watanabe is with the Department of Computer and Information Sciences, Tokyo University of Agriculture and Technology, 2-24-16, Higashikoganeishi, Tokyo, 184-8588 Japan, e-mail:shunwata@cc.tuat.ac.jp.}%
}

\begin{document}
\flushbottom
\maketitle

\begin{abstract} 
The problem of distributed data compression for function computation is
considered, where (i) the function to be computed is not necessarily
symbol-wise function and (ii) the information source has memory and may
not be stationary nor ergodic. We introduce the class of smooth
sources and give a sufficient condition on functions so that the
achievable rate region for computing coincides with the Slepian-Wolf
region (i.e., the rate region for reproducing the entire source) for any
smooth sources.  Moreover, for symbol-wise functions, the
necessary and sufficient condition for the coincidence is established.
Our result for the full side-information case is a generalization of the result by 
Ahlswede and Csisz\'ar  to sources with memory; 
our dichotomy theorem is different from Han and Kobayashi's dichotomy theorem, 
which reveals an effect of memory in distributed function computation.  
All results are given not only for fixed-length coding but also for
variable-length coding in a unified manner.  Furthermore, for the full
side-information case, the error probability in the moderate deviation
regime is also investigated.
\end{abstract}


%===============================================================================
\input{intro}%Introduction
\input{problem}%Notation & Problem
\input{theorems}%Coding Theorems
\input{proof}%Proof of Theorems
\input{conclusion}%Concluding Remarks

\appendices
\input{appendix_proof_lemma}%Proof of Lemmas

\subsection*{Acknowledgements} 
The authors would like to thank Young-Han Kim for a valuable comment on
Remark \ref{remark:mixing}.

\bibliographystyle{IEEETran}
%\bibliography{../../reference}
\bibliography{../reference}
\end{document}

%% file: intro.tex
\section{Introduction}

\begin{figure}[ht]
\centering{
\begin{minipage}{.3\textwidth}
\includegraphics[width=\textwidth]{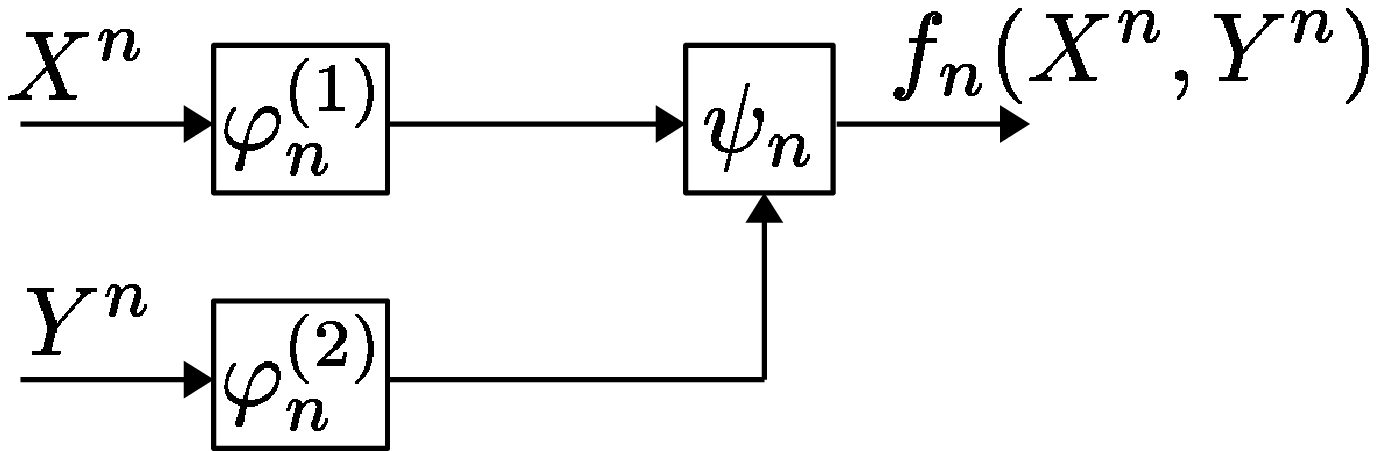}
\caption{Distributed computing}
\label{Fig:1}
\end{minipage}
\qquad
\begin{minipage}{.3\textwidth}
\includegraphics[width=\textwidth]{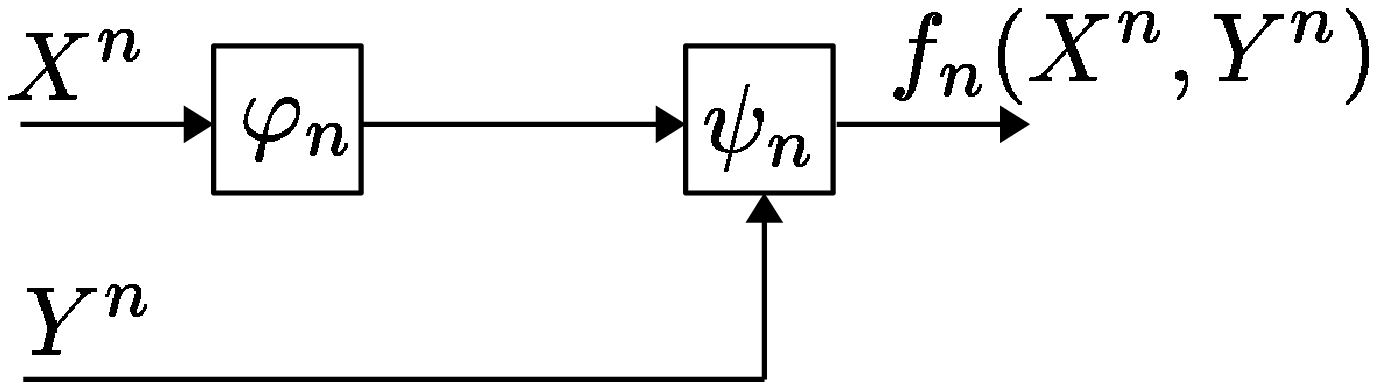}
\caption{Distributed computing with full-side-information}
\label{Fig:2}
\end{minipage}
}
\end{figure}

We study the problem of distributed data compression for function computation described in 
Fig.~\ref{Fig:1} and Fig.~\ref{Fig:2}, where 
the function to be computed is not necessarily symbol-wise function. 
In \cite{KornerMarton79}, K\"{o}rner and Marton revealed that the achievable rate region for computing modulo-sum
is strictly larger than the rate region that can be achieved by first applying 
Slepian-Wolf coding \cite{SlepianWolf73} and then computing the function.\footnote{More precisely, the modulo-sum function is a 
sensitive function explained later, and the individual rates cannot be improved from the Slepian-Wolf coding rates. In fact,
K\"{o}rner and Marton revealed that the sum rate can be improved from the Slepian-Wolf coding sum rate.}
Since then, distributed coding schemes that are tailored for some classes of functions were 
studied (e.g., see \cite[Chapter 21]{ElGamalKim}). These results are the cases such that the structure of functions can be utilized 
for distributed coding. However, not all functions have such nice structures, and even for some classes of functions,  
it is known that the Slepian-Wolf region cannot be improved at all \cite{AhlswedeCsiszar81, HanKobayashi87},
i.e., reproducing function value is as difficult as reproducing the entire source.
Thus, it is important to understand what makes distributed computation difficult,  
which is the main theme of this paper. This direction of research has been studied for i.i.d.~sources,
which will be reviewed next.

In \cite{AhlswedeCsiszar81},
Ahlswede and Csisz\'{a}r investigated distributed coding for function computation
when the full side-information is available at the decoder (see Fig.~\ref{Fig:2}); they introduced the concept of {\em sensitive functions},
and showed that the achievable rate for computing sensitive functions coincides with the achievable rate of
Slepian-Wolf coding (with full side-information) 
provided that the source is an i.i.d. source satisfying the positivity condition.\footnote{They also introduced the concept of
{\em highly sensitive functions} and showed the same result under a slightly weaker condition on the source.}
Surprisingly, the class of sensitive functions includes a
function such that the image size is just one bit. Later, El Gamal gave a simple proof of Ahlswede and Csisz\'{a}r's 
result \cite{ElGamal83}.

In \cite{HanKobayashi87}, Han and Kobayashi investigated distributed coding for function computation
with two-encoders case (see Fig.~\ref{Fig:1}); they considered the class of symbol-wise functions, and derived 
the necessary and sufficient condition of functions such that the achievable rate region coincides 
with that of Slepian-Wolf coding for any i.i.d. sources satisfying the positivity condition. In the rest of the paper,
we shall call functions satisfying Han and Kobayashi's condition {\em HK functions}.

For the class of i.i.d. sources satisfying the positivity condition, 
the above mentioned two results \cite{AhlswedeCsiszar81,HanKobayashi87} showed some classes of functions
that are difficult to compute via distributed coding. 
Then, a natural question is: 
\begin{description}
\item[($\spadesuit$)] Are functions in those classes difficult to compute  
even for wider classes of sources that have memory and may not be stationary nor ergodic?
\end{description}
In order to answer this question in a unified manner, we study distributed computation problem by information-spectral 
approach \cite{HanVerdu93, Han-spectrum}. Our contributions are summarized as follows.

%Slepian and Wolf coding \cite{SlepianWolf73} and SW region for general
%sources \cite{MiyakeKanaya95}; Variable-length SW coding for general
%sources \cite{VLSW}

\subsection{Contributions}

%In order to answer the questions posed above, we need to generalize the previously 
%studied problems \cite{AhlswedeCsiszar81,HanKobayashi87} in two directions:
%generalization with respect to sources and generalization with respect to functions.
First, we introduce a class of sources which we called {\em smooth sources};\footnote{We may call this class ``stable", but we avoid to use ``stable" since it is sometimes used to describe another concept in probability theory (eg.~\cite{Billingsley2}). In an earlier version of this paper, we also called this class ``slowly varying", but we decided to call it ``smooth" since it describes the property of the sources more accurately.}
other than the smooth condition, we do not impose any condition on sources, i.e.,
we consider general sources. 
Roughly speaking, the smooth condition says that the probability 
of a sequence does not change significantly when we flip a symbol of the sequence.
When we restrict sources to be i.i.d., then the smooth condition coincides with 
the positivity condition studied in \cite{AhlswedeCsiszar81,HanKobayashi87}. 
However, the class of smooth sources is much wider than the 
class of i.i.d. sources satisfying the positivity condition. In fact, it includes 
Markov sources with positive transition matrices or mixtures of i.i.d. sources satisfying 
positivity condition.

\begin{figure}[ht]
\centering{
\begin{minipage}{.3\textwidth}
\includegraphics[width=\textwidth]{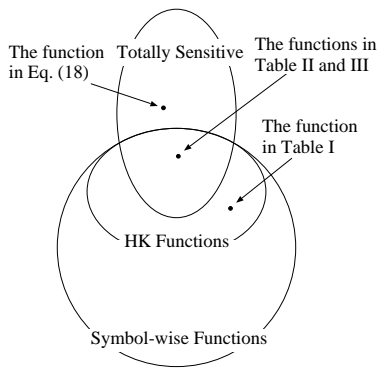}
\caption{The inclusive relation among the classes of functions.}
\label{Fig:3}
\end{minipage}
}
\end{figure}

Next, we introduce the concept of  
{\em joint sensitivity}; a function $f_n$ is said to be  jointly sensitive if $f_n(\bx,\by) \neq f_n(\hat{\bx},\hat{\by})$
whenever $\bx \neq \hat{\bx}$ and $\by \neq \hat{\by}$. Then, we introduce the class of {\em totally sensitive}
functions as the set of all functions that are sensitive in the sense 
of \cite{AhlswedeCsiszar81} and also jointly sensitive. When we restrict functions to be symbol-wise, the
class of totally sensitive functions is a strict subset of the class of HK functions. However, totally sensitive functions are not
necessarily symbol-wise. The inclusive relation among the classes of functions is summarized in Fig.~\ref{Fig:3}. 

When the full side-information is available at the decoder, we show that the Slepian-Wolf rate cannot
be improved if the function is sensitive and the source is smooth. This result generalizes the result in
\cite{AhlswedeCsiszar81} for smooth sources. Thus, for the class of sensitive functions, the answer to 
Question ($\spadesuit$) is {\em positive} in the sense that the Slepian-Wolf rate cannot be improved.

For the two-encoders case, we show that the Slepian-Wolf region cannot be improved 
if the function is totally sensitive and the source is smooth. Furthermore, for symbol-wise functions,
we show that 
the achievable region coincides with the Slepian-Wolf region for any smooth sources
if and only if the function is totally sensitive. 
In fact, for a function that satisfies Han and Kobayashi's condition but is not totally sensitive, there exists a finite state source, 
which is smooth, such that the Slepian-Wolf region can be improved. 
This dichotomy theorem can be regarded as a 
smooth source counterpart of Han and Kobayashi's dichotomy theorem \cite{HanKobayashi87}; we need the
condition that is more strict than Han and Kobayashi's condition because we broaden the class of sources.\footnote{In other words,
neither of our dichotomy theorem nor Han and Kobayashi's dichotomy theorem imply each other.}
Consequently, for the class of HK functions, the answer to Question ($\spadesuit$) is {\em negative}
in the sense that the Slepian-Wolf region can be improved; but we can say that totally sensitive functions are
difficult to compute via distributed coding for any smooth sources.

When a function is sensitive but not totally sensitive, the sum rate can be improved in general. 
To derive an outer bound for such a case, we introduce another class of functions, which we call
$r$-totally sensitive. Then, we show that the improvement of sum rate is at most $r$. 
Furthermore, we also show that there exist a smooth source and an $r$-totally sensitive function
such that our outer bound is saturated, which means that our outer bound cannot be improved anymore only from
the two assumptions: smooth condition and $r$-total sensitivity.

We also derive the following refinements of the above results.
So far, the study of distributed computing has been restricted to the fixed-length coding in 
the literature \cite{AhlswedeCsiszar81,HanKobayashi87}.
In this paper, by using the techniques developed by the authors in \cite{VLSW}, 
we show that the above mentioned results also hold even for the variable-length coding.
Furthermore, for the full side-information case, we show that the Slepian-Wolf rate cannot be
improved even in the moderate deviation regime \cite{HeMontanoYangJagmohanChen09, AltWag14}. 

Although our main contributions of this paper are structural connections between the achievable 
rate regions (or rates) for function computing and the Slepian-Wolf regions (or rates),
as a byproduct, 
we can derive explicit forms of the achievable regions (or rates) by using the corresponding 
results on the Slepian-Wolf regions (or rates).   
It is also known that distributed computing can be regarded as a special case of
distributed lossy coding studied by Yamamoto \cite{Yamamoto82} (see also \cite{OrlitskyRoche01}). Thus, our results 
may be interesting from the view point of distributed lossy coding for smooth sources.   

From technical perspective, we elaborate El Gamal's argument \cite{ElGamal83} so that it can be 
used for the wider class of sources; Lemma \ref{lemma:core} is the core of the proofs, and it enables us
to prove our main results for both fixed-length coding and variable-length coding in a unified manner.
The bounds in Lemma \ref{lemma:core} is also tight enough to be used for the moderate deviation analysis.

%\subsection{Related Works}
%lossless \cite{OrlitskyRoche01}, lossy
%\cite{Yamamoto82,FengEffrosSavari_allerton}
%\cite{DoshiShahMedardEffros10}

\subsection{Organization of Paper}

In Section \ref{sec:problem}, we introduce the coding problem investigated in this paper,
and also introduce classes of functions and classes of sources.
Then, in Section \ref{sec:theorems}, main coding theorems are stated.
The proofs of main results are given in Section \ref{sec:proof}, where proofs of some
lemmas are shown in Appendices.
%Concluding remarks ... Section \ref{sec:conclusion}

\subsection{Notation}
Throughout this paper, random variables (e.g., $X$) and their
realizations (e.g., $x$) are denoted by capital and lower case letters
respectively.  All random variables take values in some finite alphabets
which are denoted by the respective calligraphic letters (e.g., $\cX$).
Similarly, $X^n\eqtri(X_1,X_2,\dots,X_n)$ and
$x^n \eqtri(x_1,x_2,\dots,x_n)$ denote, respectively, a random vector and
its realization in the $n$th Cartesian product $\cX^n$ of $\cX$.  We
will use bold lower letters to represent vectors if the length $n$ is
apparent from the context; e.g., we use $\bx$ instead of $x^n$.

For a finite set $\cS$, $\abs{\cS}$ denotes the cardinality of $\cS$ and
$\cS^*$ denotes the set of all finite strings drawn from $\cS$. 
 For a sequence $\bms\in\cS^*$, $\abs{\bms}$ denotes the length of $\bms$.
The Hamming distance between two sequences $\bms,\hat\bms\in\cS^n$ is
defined as $d(\bms,\hat\bms)\eqtri\abs{\{i:s_i\neq\hat{s}_i\}}$.
$\cS^c$ denotes the complement of $\cS$.

Information-theoretic quantities are denoted in the usual manner
\cite{Cover2,CsiszarKorner}.  For example, $H(X|Y)$ denotes the
conditional entropy of $X$ given $Y$.  All logarithms are with respect
to base 2.

Moreover, we will use quantities defined by using the
information-spectrum method \cite{Han-spectrum}.  Here, we recall the
probabilistic limit operation: For a sequence
$\bmZ\eqtri\{Z_n\}_{n=1}^\infty$ of real-valued random variables, the
\emph{limit superior in probability} of $\bmZ$ is defined as
\begin{align}
 \plimsup_{n\to\infty}Z_n&\eqtri\inf\left\{\alpha:\lim_{n\to\infty}\Pr\{Z_n>\alpha\}=0\right\}.
\end{align}

%% file: problem.tex
\section{Problem}\label{sec:problem}
\subsection{General Setting}
Let $(\bX,\bY)=\{(X^n,Y^n)\}_{n=1}^\infty$ be a general correlated
source with finite alphabets $\cX$ and $\cY$.  We consider a sequence
$\bmf=\{f_n\}_{n=1}^\infty$ of functions
$f_n\colon\cX^n\times\cY^n\to\cZ_n$.  A \emph{variable-length code} $\code$ for computing $f_n$
is defined by a triplet $(\varphi_n^{(1)},\varphi_n^{(2)},\psi_n)$ of
the first encoder $\varphi_n^{(1)}\colon\cX^n\to\{0,1\}^*$, the second
encoder $\varphi_n^{(2)}\colon\cY^n\to\{0,1\}^*$, and a decoder
$\psi_n\colon\cC_n^{(1)}\times\cC_n^{(2)}\to\cZ_n$, where
$\cC_n^{(1)}\eqtri\{\varphi_n^{(1)}(\bx):\bx\in\cX^n\}\subseteq\{0,1\}^*$
and
$\cC_n^{(2)}\eqtri\{\varphi_n^{(2)}(\by):\by\in\cY^n\}\subseteq\{0,1\}^*$.
We assume that both of $\cC_n^{(1)}$ and $\cC_n^{(2)}$ satisfy the
prefix condition.

For each $i=1,2$, $\varphi_n^{(i)}$ is said to be a fixed-length encoder if
$\cC_n^{(i)}$ consists of codewords of the same length. A code $\code_n$
is called a \emph{fixed-length code} if both of $\varphi_n^{(i)}$
($i=1,2$) are fixed-length encoders.  Clearly, the class of all
variable-length codes includes that of all fixed-length codes as a
strict subclass.

The \emph{average codeword length} and the \emph{error probability} of
$\code_n$ are
respectively defined as
\begin{align}
\Ex\left[\abs{\varphi_n^{(1)}(X^n)}\right]&\eqtri\sum_{\bx}P_{X^n}(\bx)\abs{\varphi_n^{(1)}(\bx)},\\
\Ex\left[\abs{\varphi_n^{(2)}(Y^n)}\right]&\eqtri\sum_{\by}P_{Y^n}(\by)\abs{\varphi_n^{(2)}(\by)},\\
\intertext{and}
\error(\code_n|f_n)&\eqtri \Pr\left\{f_n(X^n,Y^n)\neq \psi_n\left(\varphi_n^{(1)}(X^n),\varphi_n^{(2)}(Y^n)\right)\right\}.
\end{align}

\begin{definition}
Given a source $(\bX,\bY)$ and a sequence of functions $\bmf=\{f_n\}_{n=1}^\infty$, 
a pair $(R_1,R_2)$ of rates is said to be \emph{achievable}, if there
exists a sequence $\{\code_n\}_{n=1}^\infty$ of codes satisfying
\begin{align}
 \lim_{n\to\infty}\error(\code_n|f_n)&=0
\label{eq1:def-achievability}
\end{align}
and
\begin{align}
 \limsup_{n\to\infty}\frac{1}{n}\Ex\left[\abs{\varphi_n^{(1)}(X^n)}\right]&\leq
 R_1,
\label{eq2:def-achievability}\\
 \limsup_{n\to\infty}\frac{1}{n}\Ex\left[\abs{\varphi_n^{(2)}(Y^n)}\right]&\leq R_2.
\label{eq3:def-achievability}
\end{align}
The set of all achievable rate pairs is denoted by $\cR^\vl(\bX,\bY|\bmf)$.
\end{definition}

\begin{definition}
Given a source $(\bX,\bY)$ and a sequence of functions $\bmf=\{f_n\}_{n=1}^\infty$, 
a pair $(R_1,R_2)$ of rates is said to be \emph{achievable by
 fixed-length coding}, if there
exists a sequence $\{\code_n\}_{n=1}^\infty$ of fixed-length codes satisfying
\eqref{eq1:def-achievability}, \eqref{eq2:def-achievability}, and \eqref{eq3:def-achievability}.
The set of all rate pairs that are achievable by fixed-length coding
is denoted by $\cR^\fl(\bX,\bY|\bmf)$.
\end{definition}

\medskip
A variable-length (resp.~fixed-length) code $\code_n$ for computing the
\emph{identity function} $f_n^\id(\bx,\by)\eqtri(\bx,\by)$ is called a
variable-length (resp.~fixed-length) \emph{Slepian-Wolf (SW) code}.

\begin{definition}[SW region]
For a source $(\bX,\bY)$, the achievable rate region 
$\cR^\vl(\bX,\bY|\bmf^\id)$ for $(\bX,\bY)$ and the sequence
$\bmf^\id\eqtri\{f_n^\id\}_{n=1}^\infty$ of identity functions is called
the \emph{Slepian-Wolf (SW) region} and denoted by 
$\cR_\SW^\vl(\bX,\bY)$.
By considering only fixed-length codes, $\cR_\SW^\fl(\bX,\bY)$ is defined similarly.
\end{definition}

\begin{remark}
From the definitions, it is apparent that $\cR_\SW^\vl(\bX,\bY)\subseteq\cR^\vl(\bX,\bY|\bmf)$ and
$\cR_\SW^\fl(\bX,\bY)\subseteq\cR^\fl(\bX,\bY|\bmf)$ for any
 $(\bX,\bY)$ and $\bmf$.
\end{remark}

\begin{remark}
A general formula for the SW region for fixed-length coding was given by
 Miyake and Kanaya \cite{MiyakeKanaya95} as
\begin{align}
 \cR_\SW^\fl(\bX,\bY)&=\left\{(R_1,R_2): R_1\geq \oH(\bX|\bY),
 R_2\geq\oH(\bY|\bX),R_1+R_2\geq \oH(\bX,\bY)\right\}
\label{eq:cR_SW_fix}
\end{align}
where
\begin{align}
 \oH(\bX,\bY)&\eqtri\plimsup_{n\to\infty}\frac{1}{n}\log\frac{1}{P_{X^nY^n}(X^nY^n)},\\
 \oH(\bX|\bY)&\eqtri\plimsup_{n\to\infty}\frac{1}{n}\log\frac{1}{P_{X^n|Y^n}(X^n|Y^n)},\\
 \oH(\bY|\bX)&\eqtri\plimsup_{n\to\infty}\frac{1}{n}\log\frac{1}{P_{Y^n|X^n}(Y^n|X^n)}.
\end{align}
As long as the authors know, a general formula for
$\cR_\SW^\vl(\bX,\bY)$ is not known. 
One of our contributions is to demonstrate that we can discuss the
 equivalence between $\cR_\SW^\vl(\bX,\bY)$ and $\cR^\vl(\bX,\bY|\bmf)$ without
 knowing the precise form of $\cR_\SW^\vl(\bX,\bY)$; for specific sources such that the precise form of
 $\cR_\SW^\vl(\bX,\bY)$ is known, we can get the precise form of $\cR^\vl(\bX,\bY|\bmf)$ as a byproduct.
\end{remark}

\medskip
As a special case of distributed computation, we are interested in the
case where $\by\in\cY^n$ is completely known at the decoder as the side-information.
We call this case as the ``full-side-information case''.
The optimal coding rates which are achievable in full-side-information
case  are defined as follows.

%\begin{definition}
%For any $(\bX,\bY)$ and $\bmf$, let
%\begin{align}
% R(\bX|\bY|\bmf)&\eqtri\inf\left\{R_1:(R_1,\log\abs{\cY})\in\cR^\vl(\bX,\bY|\bmf)\right\}.
%\end{align}
%Similarly, let
%\begin{align}
% R_\SW(\bX|\bY)&\eqtri\inf\left\{R_1:(R_1,\log\abs{\cY})\in\cR_\SW^\vl(\bX,\bY)\right\}.
%\end{align}
%Further, by considering only fixed-length codes, 
%$R^\fl(\bX|\bY|\bmf)$ and $R_\SW^\fl(\bX|\bY)$ are defined similarly.
%\end{definition}
\begin{definition}[SW rate]
For any $(\bX,\bY)$ and $\bmf$, let
\begin{align}
 R^\vl(\bX|\bY|\bmf)&\eqtri\inf\left\{R_1:(R_1,\log\abs{\cY})\in\cR^\vl(\bX,\bY|\bmf)\right\},\\
 R^\fl(\bX|\bY|\bmf)&\eqtri\inf\left\{R_1:(R_1,\log\abs{\cY})\in\cR^\fl(\bX,\bY|\bmf)\right\}.
\end{align}
Similarly, for any $(\bX,\bY)$, let
\begin{align}
 R_\SW^\vl(\bX|\bY)&\eqtri\inf\left\{R_1:(R_1,\log\abs{\cY})\in\cR_\SW^\vl(\bX,\bY)\right\},\\
 R_\SW^\fl(\bX|\bY)&\eqtri\inf\left\{R_1:(R_1,\log\abs{\cY})\in\cR_\SW^\fl(\bX,\bY)\right\}.
\end{align}
\end{definition}

\begin{remark}
From \eqref{eq:cR_SW_fix}, we have $R_\SW^\fl(\bX|\bY)=\oH(\bX|\bY)$.
A general formula for $R_\SW^\vl(\bX|\bY)$ is recently given by the authors \cite{VLSW}.
\end{remark}

%%%%%%%%%%%%%%%%%%%%%%%%%%%%%%%%%%%%%%%%%%%%%%%%%%%%%%%
\subsection{Function Classes}
In this subsection, we introduce important classes of functions
investigated in this paper.
First, we state the concept of sensitivity introduced in
\cite{AhlswedeCsiszar81} and related properties.

\begin{definition}
[Sensitivity]\label{def:sensitivity}
A function $f_n\colon\cX^n\times\cY^n\to\cZ_n$ is said to be
\emph{sensitive} conditioned on $\cY^n$ if it satisfies the following
property: If $\bx,\hbx,\by$ satisfy $f_n(\bx,\by)=f_n(\hbx,\by)$ and
$x_i\neq\hx_i$ for some $i$ then there exists $\hby\in\cY^n$ such that
$\hy_i\neq y_i$, $\hy_j=y_j$ for any $j\neq i$ and $f_n(\bx,\hby)\neq f_n(\hbx,\hby)$.
\end{definition}

\medskip
Similarly, a function $f_n\colon\cX^n\times\cY^n\to\cZ_n$ is said to be
sensitive conditioned on $\cX^n$ if it satisfies the property, where
the role of 
$\bx$ (resp.~$\hbx$) in Definition
\ref{def:sensitivity} is switched with that of $\by$ (resp.~$\hby$).

\begin{remark}
In \cite{ElGamal83}, the concenpt of $\alpha$-sensitive functions, which
includes sensitive functions as a special case, is introduced, and it is
shown that the result of \cite{AhlswedeCsiszar81}, which is proved for
sensitive functions, can be proved also for $\alpha$-sensitive
functions.
Although our results for sensitive functions hold also for 
$\alpha$-sensitive
functions, we consider only sensitive functions for simplicity.
\end{remark}

\medskip
Now, we introduce some new sensitivity conditions.

\begin{definition}
[Joint sensitivity]\label{def:joint-sensitivity}
A function $f_n\colon\cX^n\times\cY^n\to\cZ_n$ is said to be
\emph{jointly sensitive} if 
$f_n(\bx,\by)\neq f_n(\hbx,\hby)$
holds for every $\bx\neq\hbx$ and $\by\neq\hby$.
\end{definition}

\begin{definition}
[Total sensitivity]\label{def:total-sensitivity}
A function $f_n\colon\cX^n\times\cY^n\to\cZ_n$ is said to be
\emph{totally sensitive} if it is jointly sensitive and sensitive conditioned on both of
 $\cX^n$ and $\cY^n$.
\end{definition}

\begin{example}
Let $P_{\bx\by}$ be the joint type of $(\bx,\by)$ \cite{CsiszarKorner};
 i.e., $P_{\bx\by}$ is a joint distribution on $\cX\times\cY$ such as
\begin{equation}
 P_{\bx\by}(a,b)\eqtri\frac{\lvert\{i:(x_i,y_i)=(a,b)\}\rvert}{n},\quad (a,b)\in\cX\times\cY.
\end{equation}
The type function $f_n(\bx,\by)\eqtri P_{\bx\by}$ is sensitive
 conditioned on both of $\cX^n$ and $\cY^n$ but is not jointly sensitive.
Hence, it is not totally sensitive.
\end{example}

\begin{example}
The function defined by
\begin{align}
f_n(\bx,\by) \eqtri \left\{
\begin{array}{ll}
(>, \bx) & \mbox{if } \bx > \by \\
(=, \bx) & \mbox{if } \bx = \by \\
(<, \by) & \mbox{if } \bx < \by
\end{array}
\right.,
\end{align}
where $>$ and $<$ are with respect to arbitrary ordering on $\cX^n = \cY^n$, is jointly sensitive but is not sensitive conditioned on $\cX^n$ (nor $\cY^n$).
On the other hand, 
\begin{equation}
f'_n(\bx,\by)\eqtri (P_{\bx\by},f_n(\bx,\by)) 
\end{equation}
 is totally sensitive.
\end{example}

\medskip
Next, we consider special classes of symbol-wise functions.
Given a function $f$ on $\cX\times\cY$, the function $f_n$ on
$\cX^n\times\cY^n$ defined as
$f_n(\bx,\by)\eqtri(f(x_1,y_1),f(x_2,y_2),\dots,f(x_n,y_n))$ is called the symbol-wise function defined by $f$.
Now, we introduce a special class of symbol-wise functions defined by
Han and Kobayashi \cite{HanKobayashi87}.

\begin{definition}[HK functions] \label{definition:HK-functions}
 A function $f_n$ is called a \emph{Han-Kobayashi (HK) function} if 
$f_n$ is a symbol-wise function defined by some $f$ such that
\begin{enumerate}
 \item \label{HK-condition1}
       for every $a_1\neq a_2$ in $\cX$, the functions $f(a_1,\cdot)$
       and $f(a_2,\cdot)$ are distinct,
 \item \label{HK-condition2}
       for every $b_1\neq b_2$ in $\cY$, the functions $f(\cdot,b_1)$
       and $f(\cdot,b_2)$ are distinct, and 
 \item \label{HK-condition3}
 $f(a_1,b_1)\neq f(a_2,b_2)$ for every $a_1\neq a_2$ and $b_1\neq b_2$.
\end{enumerate}
\end{definition}

\medskip 
By definitions, it is easy to see that (i) an HK function is sensitive
conditioned on both of $\cX^n$ and $\cY^n$, but (ii) there exists an HK
function which is not jointly sensitive (and thus not totally
sensitive). On the other hand, it is necessary for a totally sensitive
function be an HK function.  Indeed, the next proposition gives the sufficient
and necessary condition for symbol-wise functions to be totally
sensitive. The proof of Proposition \ref{prop:HK-to-be-joint} is given
in Appendix \ref{appendix:proof_prop_HK-to-be-joint}.

\begin{proposition}
\label{prop:HK-to-be-joint}
Let $f$ be given and $f_n$ be the symbol-wise function defined by $f$.
Then $f_n$ ($n\geq 2$) is totally sensitive if and only if 
$f$ is an HK function satisfying at least
one of the following two properties:
\begin{enumerate}
 \item for all $x\in\cX$, if $f(x,y)= f(x,\hy)$ then $y=\hy$, or
 \item for all $y\in\cY$, if $f(x,y)= f(\hx,y)$ then $x=\hx$.
\end{enumerate}
\end{proposition}

\begin{example}
The function shown in Table \ref{table:HK-counter-example} is an HK
 function, but it
does not
satisfy 1) nor 2) of Proposition \ref{prop:HK-to-be-joint}.
 Thus, any $f_n$ defined by $f$ is not jointly sensitive nor totally sensitive.
Indeed, let $x^2=(0,1)$, $y^n=(0,1)$, $\hx^2=(1,1)$, and $\hy^2=(0,2)$,
 then we have $f_2(x^2,y^2)=f_2(\hx^2,\hy^2)=(0,3)$ even though $x^2\neq\hx^2$ and $y^2\neq\hy^2$.
The function shown in
Table \ref{table:HK-joint-example1} (resp.~Table
 \ref{table:HK-joint-example2}) is an HK function and satisfies 1)
 (resp. 2)) of Proposition \ref{prop:HK-to-be-joint}.
Hence, the symbol-wise function $f_n$ defined by $f$ in 
Tables \ref{table:HK-joint-example1} or \ref{table:HK-joint-example2} is totally sensitive.

\begin{table}[ht]
\begin{minipage}{.3\textwidth}
\centering{
 \caption{$f(x,y)$}\label{table:HK-counter-example}
\begin{tabular}{c|ccc}
$x\setminus y$ & 0& 1& 2\\ \hline
0 & 0 & 1 & 2 \\
1 & 0 & 3 & 3
\end{tabular}
}
\end{minipage}
~
\begin{minipage}{.3\textwidth}
\centering{
 \caption{$f(x,y)$}\label{table:HK-joint-example1}
\begin{tabular}{c|ccc}
$x\setminus y$ & 0& 1& 2\\ \hline
0 & 0 & 1 & 2 \\
1 & 0 & 3 & 4
\end{tabular}
}
\end{minipage}
~
\begin{minipage}{.3\textwidth}
\centering{
 \caption{$f(x,y)$}\label{table:HK-joint-example2}
\begin{tabular}{c|ccc}
$x\setminus y$ & 0& 1& 2\\ \hline
0 & 0 & 1 & 2 \\
1 & 3 & 3 & 3
\end{tabular}
}
\end{minipage}
\end{table}
\end{example}

%The following is an example of totally sensitive function that is not symbol-wise.
%\begin{example}
%The function defined by
%\begin{align}
%f_n(\bx,\by) \eqtri \left\{
%\begin{array}{ll}
%(P_{\bx\by}, >, \bx) & \mbox{if } \bx > \by \\
%(P_{\bx\by}, =, \bx) & \mbox{if } \bx = \by \\
%(P_{\bx\by}, <, \by) & \mbox{if } \bx < \by
%\end{array}
%\right.
%\end{align}
%is totally sensitive, where $>$ and $<$ are with respect to arbitrary ordering on $\cX^n = \cY^n$.
%\end{example}

\begin{remark}
In this subsection, several properties of functions on $\cX^n\times\cY^n$ are introduced.
In the following, we say a sequence $\bmf=\{f_n\}_{n=1}^\infty$ of functions satisfies some property, 
if $f_n$ satisfies that property for all $n=1,2,\dots$; e.g., we say ``$\bmf$ is totally sensitive'' meaning ``$f_n$ is totally sensitive for all $n=1,2,\dots$''.
\end{remark}

%%%%%%%%%%%%%%%%%%%%%%%%%%%%%%%%%%%%%%%%%%%%%%%%%%%%
\subsection{Classes of General Sources}
In this subsection, we introduce the concept of smooth sources.

\begin{definition}
\label{def:slowly-varying}
A general source $(\bX,\bY)$ is said to be \emph{smooth} with respect to $\bY$ if
there  exists a constant $0<\svc<1$, which does not depend on $n$,
satisfying
\begin{equation}
 P_{X^nY^n}(\bx,\hby)\geq \svc P_{X^nY^n}(\bx,\by)
\label{eq:svc}
\end{equation}
for any $\bx\in\cX^n$ and any $\by,\hby$ such that $d(\by,\hby)=1$. 
\end{definition}

\medskip
The definition
implies that, for a smooth source with respect to $\bY$, the
probability of joint sequences $(\bx,\by)$ does not drastically change
even if a symbol of $\by$ is replaced with another symbol.
%The constant $\svc$ is called as the varying constant.

\begin{example}[General Source with Positive Side-Information Channel]
If $Q(y|x)>q$ for all $(x,y)\in\cX\times\cY$ and 
\begin{equation}
 P_{X^nY^n}(\bx,\by)=P_{X^n}(\bx)\prod_{i=1}^nQ(y_i|x_i)
\end{equation}
then $(\bX,\bY)$ is smooth with respect to $\bY$.
\end{example}

\medskip
Similarly, a source is said to be smooth with respect to $\bX$ if
it satisfies the property, where the role of $\bx$ in Definition
\ref{def:slowly-varying} is switched with that of $\by$.  
If a source is smooth with respect to both $\bX$ and $\bY$ then we
just call it a smooth source.

As shown in the following proposition, the smooth property is
identical with the positivity condition when we consider only i.i.d.~sources.

\begin{proposition} \label{proposition:iid-positivity}
 Let $(\bX,\bY)$ be an i.i.d.~source with the joint distribution
 $P_{X_1Y_1}=P_{XY}$.
Then, $(\bX,\bY)$ is smooth if and only if $P_{XY}$ satisfies
 the positivity condition $P_{XY}(a,b)>0$ ($(a,b)\in\cX\times\cY$).
\end{proposition}

\medskip
On the other hand, as shown in following examples, the class of
smooth sources includes not only i.i.d.~sources but also Markov
sources and mixed sources.

\begin{example}[Markov Source]
Let $(\bX,\bY)$ be the source induced by a positive transition matrix
 $W(x,y|\hx,\hy)$ and a positive initial distribution
 $P_{X_1Y_1}(x,y)$. Then, by setting
\begin{align}
 \svc_1&\eqtri \min_{(x_1,y_1),(x_2,y_2),(x_3,y_3)}W(x_3,y_3|x_2,y_2)W(x_2,y_2|x_1,y_1),\\
 \svc_2&\eqtri \min_{(x_1,y_1),(x_2,y_2)}W(x_2,y_2|x_1,y_1)P_{X_1Y_1}(x_1,y_1),
\end{align}
we can find that $(\bX,\bY)$ is a smooth source with the constant $q\eqtri \min\{\svc_1,\svc_2\}$.
\end{example}

\begin{example}[Mixed Source] \label{example:mixed}
Let $(\bX_i,\bY_i)$ be a smooth source with the constant
 $\svc_i$ ($i=1,2,\dots,k$)
and consider a mixture $(\bX,\bY)$ of them such that
\begin{align}
 P_{X^nY^n}(\bx,\by)&=\sum_{i=1}^k\alpha_k P_{X_i^nY_i^n}(\bx,\by),\quad(\bx,\by)\in\cX^n\times\cY^n
\end{align}
where $\alpha_i>0$ for all $i=1,\dots,k$ and $\sum_i\alpha_i=1$.
Then, $(\bX,\bY)$ is also a smooth with the constant $\svc\eqtri\min\svc_i$.
\end{example}

\begin{remark} \label{remark:mixing}
The condition of smooth sources is different from the mixing condition that is often used as a regularity condition
for the central limit theorem in the probability theory (cf.~\cite{Billingsley2}). In fact, as we can find from 
Example \ref{example:mixed}, the class of smooth sources includes non-ergodic sources, 
which do not satisfy the mixing condition. On the other hand, an i.i.d. source that has zero probability for some
symbol is not included the class of smooth sources (cf.~Proposition \ref{proposition:iid-positivity}).  Thus, 
neither of the conditions imply each other.
\end{remark}

%% file: theorems.tex
\section{Coding Theorems}\label{sec:theorems}

\subsection{Two Encoders Case}\label{subsec:theorems_two_encs}
Our first result shows that, given a code 
$\{\code_n\}_{n=1}^\infty$ for computing a totally sensitive function
$\bmf$, we can construct a SW code $\{\hat\code_n\}_{n=1}^\infty$ such
that the coding rates of $\{\hat\code_n\}_{n=1}^\infty$ are asymptotically same as 
$\{\code_n\}_{n=1}^\infty$ and the error probability of
$\{\hat\code_n\}_{n=1}^\infty$ is vanishing as $n\to\infty$, provided
that $(\bX,\bY)$ is smooth.

\begin{theorem}
\label{thm:two_enc_func_to_SW}
Suppose that $(\bX,\bY)$ is smooth and $\bmf$ is totally sensitive. Then,
for any variable-length (resp.~fixed-length) code $\{\code_n\}_{n=1}^\infty$ for computing $\bmf$
satisfying
\eqref{eq1:def-achievability}--\eqref{eq3:def-achievability},  there exists
a variable-length (resp.~fixed-length) SW code
 $\{\hat\code_n\}_{n=1}^\infty=\{(\hvarphi_n^{(1)},\hvarphi_n^{(2)},\hpsi_n)\}_{n=1}^\infty$ such that 
\begin{align} \label{eq:theorem1-1}
 \lim_{n\to\infty}\error(\hat\code_n|f_n^\id)&=0
\end{align}
and
\begin{align}
 \limsup_{n\to\infty}\frac{1}{n}\Ex\left[\abs{\hvarphi_n^{(1)}(X^n)}\right]&\leq
 R_1, \label{eq:theorem1-2} \\
 \limsup_{n\to\infty}\frac{1}{n}\Ex\left[\abs{\hvarphi_n^{(2)}(Y^n)}\right]&\leq R_2.
 \label{eq:theorem1-3}
\end{align}
\end{theorem}

\medskip
The proof will be given in the next section.
As a consequence of Theorem \ref{thm:two_enc_func_to_SW}, we have the
following theorem, which shows 
that the achievable rate region for a smooth source $(\bX,\bY)$ and a totally sensitive
function $\bmf$ is identical with the SW region.

\begin{theorem}
\label{thm:two_enc_general}
Suppose that $(\bX,\bY)$ is smooth and $\bmf$ is totally sensitive. Then we
have
\begin{align}
 \cR^\fl(\bX,\bY|\bmf)&=\cR_{\SW}^\fl(\bX,\bY)
\intertext{and}
 \cR^\vl(\bX,\bY|\bmf)&=\cR_{\SW}^\vl(\bX,\bY).
\end{align} 
\end{theorem}

\medskip
Theorem \ref{thm:two_enc_general} states that the total sensitivity is 
a sufficient condition for the set of all achievable rates to coincide with the SW region.
It should be noted that total sensitivity is not necessary; See 
Remark \ref{remark:totally_sensitivity_is_not_necessary}
below for more details.

On the other hand, if we restrict our attention to the 
class of symbol-wise functions, we can also prove the converse statement, i.e.,
the total sensitivity is the
necessary and sufficient condition for the set of all achievable rates
to coincide with the SW region.
More precisely, we have the following theorem.

\begin{theorem}
\label{thm:two_enc_symbolwise_fix}
  Let $\bmf$ be a sequence of symbol-wise functions.
Then $\cR^\fl(\bX,\bY|\bmf)=\cR_{\SW}^\fl(\bX,\bY)$ for all smooth
 sources $(\bX,\bY)$ if and only if $\bmf$ is totally sensitive. 
\end{theorem}

\medskip
Now, let us compare our result with that of Han and Kobayashi \cite{HanKobayashi87}.

\begin{proposition}
[Theorem 1 of \cite{HanKobayashi87}]\label{prop:HanKobayashi}
Let $\bmf$ be a sequence of symbol-wise functions.  Then
$\cR^\fl(\bX,\bY|\bmf)=\cR_{\SW}^\fl(\bX,\bY)$ for all i.i.d.~sources
$(\bX,\bY)$ satisfying the positivity condition $P_{X_1Y_1}(x,y)>0$
 if and only if $\bmf$ is an HK function.
\end{proposition}

\medskip
Comparison of Theorem \ref{thm:two_enc_symbolwise_fix} with Proposition
\ref{prop:HanKobayashi} implies that the condition given by Han and Kobayashi
\cite{HanKobayashi87} is no longer sufficient for
$\cR^\fl(\bX,\bY|\bmf)=\cR_{\SW}^\fl(\bX,\bY)$, when we consider not
only i.i.d.~sources but also sources with memory.\footnote{Note that neither Theorem
\ref{thm:two_enc_symbolwise_fix} nor Proposition \ref{prop:HanKobayashi}
subsumes the other.}

Further, we can generalize the result for the variable-length coding case.

\begin{theorem}
\label{thm:two_enc_symbolwise_var}
Let $\bmf$ be a sequence of symbol-wise functions.
Then $\cR^\vl(\bX,\bY|\bmf)=\cR_\SW^\vl(\bX,\bY)$ for all smooth
 sources $(\bX,\bY)$ if and only if $\bmf$ is totally sensitive. 
\end{theorem}

%%%%%%%%%%%%%%%%%%%%%%%%%%%%%%%%%%%%%%%%%%%%%%%%%%%%%%%
\subsection{Full-Side-Information Case}\label{subsec:theorems_full_side_inf}

Theorem \ref{thm:two_enc_general} assumes 
 the smooth property of the source and 
the total sensitivity of functions.
In the full-side-information case, weaker conditions are sufficient to show the corresponding result.
Indeed we have the following theorem.

\begin{theorem}
\label{thm:fullside_fix}
 Suppose that $(\bX,\bY)$ is smooth with respect to $\bY$ and $\bmf$ is
sensitive conditioned on $\cY^n$. Then we have
\begin{align}
 R^\fl(\bX|\bY|\bmf)=R_\SW^\fl(\bX|\bY).
\end{align}
\end{theorem}

\medskip
As a corollary of the theorem, we can derive the first half of Theorem 3 of \cite{AhlswedeCsiszar81}.

\begin{corollary}
[\cite{AhlswedeCsiszar81}]\label{corollary:AC81}
Suppose that $(\bX,\bY)$ is an i.i.d.~source satisfying the positivity
condition $P_{X_1Y_1}(x,y)>0$ and $\bmf$ is
sensitive conditioned on $\cY^n$. Then we have
\begin{align}
 R^\fl(\bX|\bY|\bmf)=R_\SW^\fl(\bX|\bY).
\end{align}
\end{corollary}

\begin{remark}
We can also derive Lemmas 1 and 2 of \cite{HanKobayashi87} by applying
Theorem \ref{thm:fullside_fix} (or Corollary \ref{corollary:AC81}) to
symbol-wise functions.
\end{remark}

\begin{remark}
\label{remark:second_half}
In the second half of Theorem 3 of \cite{AhlswedeCsiszar81}, it is shown
 that if $\bmf$ is highly sensitive then 
Corollary \ref{corollary:AC81} holds even under the weaker condition.
% that for every $a_1\neq a_2$ in $\cX^n$ the number of elements
% $b\in\cY$ with
% \begin{align}
%  P_{X_1Y_1}(a_1,b)\cdot P_{X_1Y_1}(a_2,b)>0
% \end{align}
% is different from one.
Similarly, we can prove that if $\bmf$ is highly sensitive then Theorem
\ref{thm:fullside_fix} holds even under the condition weaker than the
smooth property, and thus, we can derive also the second half of Theorem 3 of
\cite{AhlswedeCsiszar81} as a corollary.  See
Section \ref{subsec:weaker_condition} for more details.
\end{remark}

\medskip
Further, we can generalize the result for the variable-length coding case.

\begin{theorem}
\label{thm:fullside_var}
 Suppose that $(\bX,\bY)$ is smooth with respect to $\bY$ and $\bmf$ is
sensitive conditioned on $\cY^n$. Then we have
\begin{align}
 R^\vl(\bX|\bY|\bmf)=R_\SW^\vl(\bX|\bY).
\end{align}
\end{theorem}

%%%%%%%%%%%%%%%%%%%%%%%%%%%%%%%%%%%%%%%%%%%%%%%%%%%%%%%
\subsection{Weaker Condition on Sources}\label{subsec:weaker_condition}
So far, we consider only smooth sources for simplicity.
In this subsection, we show
that all our results in Sections \ref{subsec:theorems_two_encs} and \ref{subsec:theorems_full_side_inf}
are true even for a class of sources wider than 
smooth sources, provided
that the function $\bmf$ is highly sensitive in the sense of \cite{AhlswedeCsiszar81}.

\begin{definition}
A function $f_n\colon\cX^n\times\cY^n\to\cZ_n$ is said to be
\emph{highly sensitive} conditioned on $\cY^n$ if for any $a_1\neq a_2$ in $\cX$ and $b_1\neq b_2$
in $\cY$ the following
property holds: 
If $\bx,\hbx,\by$ satisfy $f_n(\bx,\by)=f_n(\hbx,\by)$,
$x_i=a_1$,
$\hx_i=a_2$, and 
$y_i=b_1$ 
for some $i$ then 
for
$\hby\in\cY^n$ obtained from $\by$ by replacing the $i$th component by 
$b_2$ we always have $f_n(\bx,\hby)\neq f_n(\hbx,\hby)$.
\end{definition}

\medskip
Similarly, the concept of ``the highly sensitivity conditioned on
$\cX^n$'' is defined. Further, by replacing the sensitivity with the
highly sensitivity in Definition \ref{def:total-sensitivity}, the \emph{highly total sensitivity} is defined.

Now, we define a class of sources which is wider than the class of
smooth sources.

\begin{definition}
\label{def:weakly-slowly-varying}
A general source $(\bX,\bY)$ is said to be \emph{weakly smooth} with respect to $\bY$ if
there  exists a constant $0<\svc<1$, which does not depend on $n$,
satisfying the following property:
For any 
$\bx\neq\hbx$ and $\by$ satisfying $P_{X^nY^n}(\bx,\by)\cdot P_{X^nY^n}(\hbx,\by)>0$, whenever
$x_i\neq\hx_i$, there exists
$\hby\in\cY^n$ such that 
$\hy_i\neq y_i$, $\hy_j=y_j$ for any $j\neq i$ and 
\begin{align}
 P_{X^nY^n}(\bx,\hby)&\geq \svc P_{X^nY^n}(\bx,\by),\label{eq1:def_weakly_slowly_varying}\\
 P_{X^nY^n}(\hbx,\hby)&\geq \svc P_{X^nY^n}(\hbx,\by).\label{eq2:def_weakly_slowly_varying}
\end{align}
\end{definition}

\medskip
Similarly, a source is said to be weakly smooth with respect to $\bX$  if it satisfies the property, where the role of
$\bX$  in Definition \ref{def:weakly-slowly-varying} is switched with that of $\bY$ . If a source is weakly smooth with respect to both $\bX$  and $\bY$  then we just call it a weakly smooth source.

Then, we can modify theorems
in Sections \ref{subsec:theorems_two_encs} and \ref{subsec:theorems_full_side_inf}
as in the following theorem.

\begin{theorem} 
\label{thm:weakly_slowly_varying}
Theorems \ref{thm:two_enc_func_to_SW}, \ref{thm:two_enc_general}, \ref{thm:two_enc_symbolwise_fix}, and
 \ref{thm:two_enc_symbolwise_var} hold even when we replace 
``smooth'' (resp.~``totally sensitive'') with 
``weakly smooth'' (resp.~``highly totally sensitive'').
Further, 
Theorems
\ref{thm:fullside_fix} and \ref{thm:fullside_var}
hold even when we replace 
``smooth'' (resp.~``sensitive'') with 
``weakly smooth'' (resp.~``highly sensitive'').
\end{theorem}

\medskip
Especially, as mentioned in Remark \ref{remark:second_half}, the second
half of Theorem 3 of \cite{AhlswedeCsiszar81} can be derived as a
corollary of the above theorem, since the following proposition holds. 
The proof of Proposition \ref{prop:AC_weak_condition} is given in Appendix \ref{appendix:proof_AC_weak_condition}.

\begin{proposition}
\label{prop:AC_weak_condition}
 Let $(\bX,\bY)$ be an i.i.d.~source with the joint distribution
 $P_{X_1Y_1}=P_{XY}$.
Then, $(\bX,\bY)$ is weakly smooth with respect to $\bY$ if and only if $P_{XY}$ satisfies
the condition  that for every $a_1\neq a_2$ in $\cX$ the number of elements
 $b\in\cY$ with
\begin{align}
 P_{XY}(a_1,b)\cdot P_{XY}(a_2,b)>0
\end{align}
is different from one.
\end{proposition}

%%%%%%%%%%%%%%%%%%%%%%%%%%%%%%%%%%%%%%%%%%%%%%%%%%%%%%%
\subsection{Weaker Condition on Functions}\label{subsec:theorems_not_totally_sensitive}

So far, we considered conditions on functions so that
$\cR^{\mathsf{**}}(\bX,\bY|\bmf)=\cR_{\SW}^{\mathsf{**}}(\bX,\bY)$
($\mathsf{**}=\fl/\vl$) holds.  As a byproduct, we can give explicit
forms of $\cR^{\mathsf{**}}(\bX,\bY|\bmf)$ by using the corresponding
results on $\cR_{\SW}^{\mathsf{**}}(\bX,\bY)$, provided that $\bmf$
satisfies conditions for the coincidence of two regions.  In this
subsection, we consider functions which does not satisfy conditions for
the coincidence.  We introduce a class of functions wider than the
totally sensitive functions, and give an outer bound on
$\cR^{\mathsf{**}}(\bX,\bY|\bmf)$ of $\bmf$ in this class.

To define a new class of functions, we introduce a notation.  Given a
function $f_n\colon\cX^n\times\cY^n\to\cZ_n$ and $z_n\in\cZ_n$, let
$\EQ(z_n|f_n)$ be the maximum number $J$ such that we can choose $J$
pairs
$(\bx_1,\by_1),(\bx_2,\by_2),\dots,(\bx_J,\by_J)\in\cX^n\times\cY^n$
satisfying $\bx_i\neq\bx_j$ and $\by_i\neq\by_j$ for all $i\neq j$ and
$z_n=f_n(\bx_1,\by_1)=f_n(\bx_2,\by_2)=\dots=f_n(\bx_J,\by_J)$.

\begin{definition}
Fix a number $r\geq 0$.
A function $f_n\colon\cX^n\times\cY^n\to\cZ_n$ is said to be
 \emph{$r$-totally sensitive} if it satisfies
\begin{equation}
 \limsup_{n\to\infty}\frac{1}{n}\log \max_{z_n\in\cZ_n}\EQ(z_n|f_n)\leq r
\label{eq:r-totally-sensitive}
\end{equation}
and sensitive conditioned on both of $\cX^n$ and $\cY^n$.
\end{definition}

\begin{remark}
 Note that the maximum $J$ such that we can choose $J$ pairs
$(\bx_1,\by_1),(\bx_2,\by_2),\dots,(\bx_J,\by_J)\in\cX^n\times\cY^n$
satisfying $\bx_i\neq\bx_j$ and $\by_i\neq\by_j$ for all $i\neq j$ is
 $\min\{\abs{\cX^n},\abs{\cY^n}\}$. Hence, the definition of 
$r$-total sensitivity is meaningless if 
 $r> \barr\eqtri \min\{\log\abs{\cX},\log\abs{\cY}\}$.
\end{remark}

\begin{theorem}
\label{thm:not_totally_sensitive_bounds}
Suppose that $(\bX,\bY)$ is smooth and $\bmf$ is $r$-totally sensitive. Then we
have
\begin{align}
\cR^{\mathsf{**}}(\bX,\bY|\bmf)
&\subseteq
\bigl\{
(R_1,R_2)
:0\leq\exists\lambda\leq 1,(R_1+\lambda
 r,R_2+(1-\lambda)r)\in\cR_{\SW}^{\mathsf{**}}(\bX,\bY),\nonumber\\
&\qquad\qquad R_1\geq R_{\SW}^{\mathsf{**}}(\bX|\bY), R_2\geq R_{\SW}^{\mathsf{**}}(\bY|\bX)
\bigr\}
\label{eq:thm_not_totally_sensitive_bounds}
\end{align} 
where $\mathsf{**}=\fl/\vl$.
\end{theorem}

\begin{figure}[ht]
\centering{
\begin{minipage}{.35\textwidth}
\includegraphics[width=\textwidth]{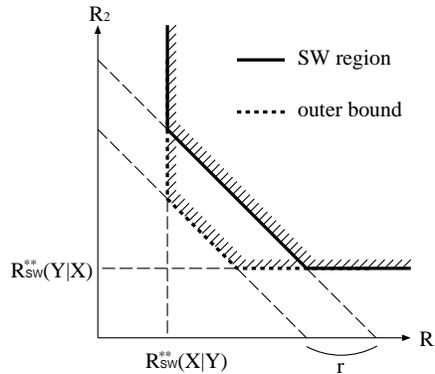}
\caption{The outer bound given in Theorem \ref{thm:not_totally_sensitive_bounds}.}
\label{Fig:4}
\end{minipage}
}
\end{figure}

\begin{remark}
\label{remark:totally_sensitivity_is_not_necessary} 
Theorem \ref{thm:not_totally_sensitive_bounds} states that only the sum rate can be improved at most $r$; see Fig.~\ref{Fig:4}.
Note that if $f_n$
is totally sensitive then $\EQ(z_n|f_n)\leq 1$ for any $z_n\in\cZ_n$ and
thus \eqref{eq:r-totally-sensitive} holds with $r=0$. In other
words, the class of totally sensitive functions can be seen as a special
case of $0$-totally sensitive functions.  Moreover, by Theorem
\ref{thm:not_totally_sensitive_bounds}, we can say that $0$-total
sensitivity is sufficient for
$\cR^{\mathsf{**}}(\bX,\bY|\bmf)=\cR_{\SW}^{\mathsf{**}}(\bX,\bY)$.  In
this sense, Theorem \ref{thm:not_totally_sensitive_bounds} is a
generalization of Theorem \ref{thm:two_enc_general}.
\end{remark}

\medskip
Moreover, as shown in the theorem below, there exist $r$-totally
sensitive function and a smooth source for which the outer bound
given in Theorem \ref{thm:not_totally_sensitive_bounds} is tight.

\begin{theorem}
\label{thm:example_outer_bound}
For any $\delta>0$ and $0\leq r\leq \barr$, there exist $r$-totally
sensitive function $\bmf$ and a smooth source $(\bX,\bY)$ such that
\begin{align}
 \cR^{\mathsf{**}}(\bX,\bY|\bmf)&\supseteq
\left\{
(R_1,R_2): R_1\geq \delta+(1-\rho)\log\abs{\cX}, R_2\geq \delta+(1-\rho)\log\abs{\cY}
\right\}
\label{eq1:thm_example_outer_bound}
\intertext{and}
 \cR_{\SW}^{\mathsf{**}}(\bX,\bY)&\subseteq\bigl\{
(R_1,R_2): R_1\geq (1-\rho)\log\abs{\cX}, R_2\geq (1-\rho)\log\abs{\cY},\nonumber\\
&\qquad\qquad R_1+R_2\geq r-\delta+(1-\rho)\log\abs{\cX}\abs{\cY}
\bigr\}
\label{eq2:thm_example_outer_bound}
\end{align}
where $\rho\eqtri r/\barr$ and $\mathsf{**}=\fl/\vl$.
\end{theorem}

\medskip

By taking $\delta \to 0$ in Theorem \ref{thm:example_outer_bound}, 
we can make the bound \eqref{eq:thm_not_totally_sensitive_bounds} arbitrarily tight.
Hence, our outer bound cannot be improved anymore only from $r$-total
sensitivity and smooth condition.

%%%%%%%%%%%%%%%%%%%%%%%%%%%%%%%%%%%%%%%%%%%%%%%%%%%%%%%%%%%%
\subsection{Moderate Deviation}\label{subsec:theorems_moderate_deviation}
In this subsection, we assume that $(\bX,\bY)$ is an i.i.d.~source with
the joint distribution $P_{X_1Y_1}=P_{XY}$, and we consider the full side-information case.
The results in Section \ref{subsec:theorems_full_side_inf} states that 
$R^{\mathsf{**}}(\bX|\bY|\bmf)=R_\SW^{\mathsf{**}}(\bX|\bY) = H(X|Y)$ ($\mathsf{**}=\fl/\vl$).
In the following, we conduct more refined analysis in the moderate deviation regime.

For real numbers $t\in(0,1/2)$ and $\gamma>0$,
and a sequence of functions $\bmf = \{f_n\}_{n=1}^\infty$, let 
\begin{align}
 e^\vl(t,\gamma|\bmf) &\eqtri \liminf_{n\to\infty} -\frac{1}{n^{1-2t}}\log\min_{\code_n}\error(\code_n|f_n)
\end{align}
where the minimum is taken over all sequences of codes
$\{ \code_n \}_{n=1}^\infty =\{ (\varphi_n^{(1)},\psi_n) \}_{n=1}^\infty$ for computing $\bmf = \{f_n\}_{n=1}^\infty$ satisfying
\begin{align} \label{eq:moderate-rate-constraint}
\limsup_{n\to\infty} \frac{1}{n^{1-t}} \left( \Ex\left[\abs{
\varphi_n^{(1)}(X^n)
}\right] - n H(X|Y) \right) \leq \gamma.
\end{align}
Similarly, by taking the minimum over all fixed-length codes, 
$e^\fl(t,\gamma|\bmf)$ is defined.
Further, by considering the identity function $f_n^\id$ and SW codes, 
$e_{\SW}^\vl(t,\gamma)$ 
and $e_{\SW}^\fl(t,\gamma)$ 
are defined. The single-letter characterization of $e_{\SW}^\vl(t,\gamma)$ and
$e_{\SW}^\fl(t,\gamma)$ are obtained by He {\em et.~al.}\cite{HeMontanoYangJagmohanChen09}. The following theorem states that 
computing sensitive function is as difficult as reproducing $\mathbf{X}$ itself even for the moderate deviation regime.

\begin{theorem} \label{thm_moderate_deviation}
Suppose that $P_{XY}$ satisfies positivity condition and $\bmf$ is sensitive. Then, we have
\begin{align}
 e^\fl(t,\gamma|\bmf) &= e_{SW}^\fl(t,\gamma), \label{eq:moderate-fl} \\
 e^\vl(t,\gamma|\bmf)&= e_{SW}^\vl(t,\gamma) \label{eq:moderate-vl}
\end{align}
for every $t \in (0,1/2)$ and $\gamma>0$.
\end{theorem}

%% file: proof.tex
\section{Proof of Theorems}\label{sec:proof}

%%%%%%%%%%%%
\subsection{Preliminaries for Proofs} 

First we introduce some notations used in this section.
$\bm{1}$ denotes the indicator function, e.g., $\bm{1}[s\in\cS]=1$ if
$s\in\cS$ and $0$ otherwise;
$h(p)$ is binary entropy function;
$[a]^+$ is $0$ if $a < 1$ and $\lfloor a \rfloor$ if $a \ge 1$. For given $0 < \beta < 1/2$, let 
\begin{align}
v_n(\beta) &\eqtri \sum_{i=1}^{\lceil n \beta \rceil -1} (|\cX|-1)^i \binom{n}{i} \\
 &\le n |\cX|^{n\beta} 2^{n h(\beta)}, \\
u_n(\beta) &\eqtri \sum_{i=1}^{\lceil n \beta \rceil -1} (|\cY|-1)^i \binom{n}{i} \\
 &\le n |\cY|^{n\beta} 2^{n h(\beta)}.
\end{align}

For a given code $\Phi_n = (\varphi_n^{(1)},\varphi_n^{(2)},\psi_n)$, we abbreviate the length 
of codewords by 
\begin{align}
\ell_n^{(1)}(\bx) &\eqtri \abs{\varphi_n^{(1)}(\bx)}, \\
\ell_n^{(2)}(\bx) &\eqtri \abs{\varphi_n^{(2)}(\bx)}.
\end{align}
Without loss of generality,\footnote{Note that for any encoder $\varphi_n^{(1)}$,
we can modify $\varphi_n^{(1)}$ without increasing the error probability and obtain an encoder
$\tilde{\varphi}_n^{(1)}$ satisfying $\abs{\tilde{\varphi}_n^{(1)}(\bx)} \le \max\left[ 1+\abs{\varphi_n^{(1)}(\bx)}, 1 + n \lceil \log |\cX| \rceil \right]$.} 
 we assume that there are $L_1 < \infty$ and $L_2 < \infty$ such that
\begin{align}
\ell_n^{(1)}(\bx) &\le n L_1,~~~\mbox{for all } \bx \in \cX^n, \\
\ell_n^{(2)}(\bx) &\le n L_2,~~~\mbox{for all } \by \in \cY^n.
\end{align}
Let 
\begin{align}
\cD_n \eqtri \left\{ (\bx,\by) : \psi_n(\varphi_n^{(1)}(\bx), \varphi_n^{(2)}(\by)) = f_n(\bx,\by) \right\}
\end{align}
be the set of all correctly decodable sequences. When we analyze the performance of a variable length code via
information spectrum approach, the following typical-like sets play an important role:
\begin{align}
\cT_{n,1} &\eqtri \left\{ (\bx,\by) : \frac{1}{n} \log \frac{1}{P_{X^n|Y^n}(\bx|\by)} \le \frac{\ell_n^{(1)}(\bx)}{n} + \delta \right\}, \\
\cT_{n,2} &\eqtri \left\{ (\bx,\by) : \frac{1}{n} \log \frac{1}{P_{Y^n|X^n}(\by|\bx)} \le \frac{\ell_n^{(2)}(\by)}{n} + \delta \right\}, \\
\cT_{n,0} &\eqtri \left\{ (\bx,\by) : \frac{1}{n} \log \frac{1}{P_{X^nY^n}(\bx,\by)} 
  \le \frac{\ell_n^{(1)}(\bx)}{n} + \frac{\ell_n^{(2)}(\by)}{n} + \delta \right\},
\end{align}
where $\delta > 0$ is any real number specified later.
The following lemma is the core of the proofs of coding theorems, 
which connect the combinatorial property, i.e., the sensitivity of a function, 
to a probabilistic analysis. The proof of the lemma will be given in Appendix \ref{appendix:proof_lemma}.
%%%%%%
\begin{lemma} \label{lemma:core}
For any code $\Phi_n$ and real numbers $\beta, \delta > 0$, 
if $f_n$ is sensitive conditioned on $\cY^n$ and $(\bX,\bY)$ is smooth with respect to $\bY$, then we have
\begin{align}
P_{X^nY^n}(\cD_n \cap \cT_{n,1}^c) &\le \frac{2 |\cY|}{\beta q} \error(\code_n|f_n) + (v_n(\beta) + 1) 2^{- n \delta}. \label{eq:core1} 
\end{align}
Similarly, if $f_n$ is sensitive conditioned on $\cX^n$
and $(\bX,\bY)$ is smooth with respect to $\bX$, then we have
\begin{align}
P_{X^nY^n}(\cD_n \cap \cT_{n,2}^c) &\le \frac{2 |\cX|}{\beta q} \error(\code_n|f_n) + (u_n(\beta) + 1) 2^{- n \delta}. \label{eq:core2} 
\end{align}
Furthermore, if $f_n$ is totally sensitive and
$(\bX,\bY)$ is smooth, then we have
\begin{align}
P_{X^nY^n}(\cD_n \cap \cT_{n,0}^c) &\le \frac{2(|\cX|+|\cY|) }{\beta q} \error(\code_n|f_n) + 
 [(v_n(\beta) +1) + (u_n(\beta)+1)] 2^{- n \delta}. \label{eq:core3}
\end{align}
\end{lemma}

The following lemma is an immediate consequence of Lemma \ref{lemma:core}.
\begin{lemma} \label{lemma:non-typical}
For any $\delta > 0$ and any code satisfying \eqref{eq1:def-achievability}, if $\bmf$ is totally sensitive, we have
\begin{align}
\lim_{n\to\infty} P_{X^nY^n}(\cT_{n,1}^c \cup \cT_{n,2}^c \cup \cT_{n,0}^c) = 0.
\end{align}
\end{lemma}
\begin{IEEEproof}
We have
\begin{align}
& P_{X^nY^n}(\cT_{n,1}^c \cup \cT_{n,2}^c \cup \cT_{n,0}^c) \nonumber\\
&= P_{X^nY^n}(\cD_n^c \cap (\cT_{n,1}^c \cup \cT_{n,2}^c \cup \cT_{n,0}^c)) 
 + P_{X^nY^n}(\cD_n \cap (\cT_{n,1}^c \cup \cT_{n,2}^c \cup \cT_{n,0}^c)) \\
&\le \error(\code_n|f_n) + P_{X^nY^n}(\cD_n \cap \cT_{n,1}^c) + P_{X^nY^n}(\cD_n \cap \cT_{n,2}^c) + P_{X^nY^n}(\cD_n \cap \cT_{n,0}^c).
\end{align}
Then, we apply Lemma \ref{lemma:core} by taking sufficiently small $\beta > 0$ so that 
$v_n(\beta) 2^{-n\delta}$ and $u_n(\beta) 2^{-n\delta}$ converges to $0$ as $n\to\infty$.
\end{IEEEproof}

%%%%%%%%%%%%

%%%%%%%%%%%%%%%%%%%%%%%%%%%%%%%%%%%%%%%%%%%%%%%%%%%%%%%%%%%
\subsection{Proof of Theorem \ref{thm:two_enc_func_to_SW}} \label{subsection:proof-thm:two_enc_func_to_SW}

For a given (variable-length) code $\Phi_n$ for function computation, we construct a SW code
by using a random binning of adaptive length.\footnote{We only show the statement for variable-length
coding since the statement for fixed-length coding can be proved as a special case of the former.}
Let 
\begin{align}
\tilde{\ell}_n^{(1)}(\bx) &\eqtri \lceil \ell_n^{(1)}(\bx) + 2n\delta \rceil, \\
\tilde{\ell}_n^{(2)}(\bx) &\eqtri \lceil \ell_n^{(2)}(\bx) + 2n\delta \rceil,
\end{align}
and for each integer $l$, let 
\begin{align}
\cS_{n,1}(l) &\eqtri \left\{ (\bx,\by) : \frac{1}{n} \log \frac{1}{P_{X^n|Y^n}(\bx|\by)} \le \frac{l}{n} - \delta \right\}, \\
\cS_{n,2}(l) &\eqtri \left\{ (\bx,\by) : \frac{1}{n} \log \frac{1}{P_{Y^n|X^n}(\by|\bx)} \le \frac{l}{n} - \delta \right\}, \\
\cS_{n,0}(l) &\eqtri \left\{ (\bx,\by) : \frac{1}{n} \log \frac{1}{P_{X^nY^n}(\bx,\by)} \le \frac{l}{n} - \delta \right\}.
\end{align}
Further, for integers $l_1$ and $l_2$, let
\begin{align}
\cS_n(l_1,l_2) \eqtri \cS_{n,1}(l_1) \cap \cS_{n,2}(l_2) \cap \cS_{n,0}(l_1+l_2).
\end{align}
Note that, for any $\by \in \cY^n$, we have
\begin{align}
|\{ \hat{\bx} : (\hat{\bx},\by) \in \cS_n(l_1,l_2) \} |
&\le | \{ \hat{\bx} : (\hat{\bx},\by) \in \cS_{n,1}(l_1) \}| \\
&\le 2^{l_1 - n \delta}.
\end{align}
Similarly, for any $\bx \in \cX^n$, we have
\begin{align}
|\{ \hat{\by} : (\bx,\hat{\by}) \in \cS_n(l_1,l_2) \}| \le 2^{l_2 - n\delta }
\end{align}
and
\begin{align}
|\{ (\hat{\bx},\hat{\by}) : (\hat{\bx},\hat{\by}) \in \cS_n(l_1,l_2) \} | \le 2^{l_1 + l_2 - n\delta}.
\end{align}

Now, we construct a SW code as follows:
\begin{itemize}
\item Given $\bx \in \cX^n$, the encoder 1
\begin{enumerate}
\item sends the integer $l_1 = \tilde{\ell}_n^{(1)}(\bx)$ by using at most $2 (\lfloor \log \tilde{\ell}_n^{(1)}(\bx) \rfloor + 1)$ 
 bits \cite{Elias75}, and then
\item by using a random bin-code with $\tilde{\ell}_n^{(1)}(\bx)$ bits, sends the bin-index $m_1$ of $\bx$.
\end{enumerate}

\item Given $\by \in \cY^n$, the encoder 2
\begin{enumerate}
\item sends the integer $l_2 = \tilde{\ell}_n^{(2)}(\by)$ by using at most $2(\lfloor \log \tilde{\ell}_n^{(1)}(\bx)  \rfloor+ 1)$ 
 bits \cite{Elias75}, and then
\item by using a random bin-code with $\tilde{\ell}_n^{(2)}(\by)$ bits, sends the bin-index $m_2$ of $\by$.
\end{enumerate}

\item The decoder
\begin{enumerate}
\item extracts $l_1$, $l_2$, $m_1$, and $m_2$ from the received codewords, and then
\item looks for the unique pair $(\bx,\by)$ such that $(\bx,\by) \in \cS_n(l_1,l_2)$, 
$\tilde{\ell}_n^{(1)}(\bx) = l_1$, $\tilde{\ell}_n^{(2)}(\by) = l_2$, and the bin-index of $\bx$ (resp.~$\by$) is $m_1$ (resp.~$m_2$).
\end{enumerate}
\end{itemize}

By using the standard argument, we can upper bound the average error probability $\mathbb{E}[\error(\hat\code_n|f_n^\id)]$
of the constructed SW code with respect to random bin-coding by
\begin{align}
\mathbb{E}[\error(\hat\code_n|f_n^\id)]
&\le \Pr\Bigg\{ \frac{1}{n} \log \frac{1}{P_{X^n|Y^n}(X^n|Y^n)} > \frac{\tilde{\ell}_n^{(1)}(X^n)}{n} - \delta \text{ or } \nonumber\\
&~~~\frac{1}{n} \log \frac{1}{P_{Y^n|X^n}(Y^n|X^n)} > \frac{\tilde{\ell}_n^{(2)}(Y^n)}{n} - \delta \text{ or } \nonumber\\
&~~~\frac{1}{n} \log \frac{1}{P_{X^nY^n}(X^n,Y^n)} > \frac{\tilde{\ell}_n^{(1)}(X^n) + \tilde{\ell}_n^{(2)}(Y^n)}{n} - \delta \Bigg\} \nonumber\\
&+ \sum_{\bx,\by} P_{X^nY^n}(\bx,\by) 
 \frac{\left| \left\{ \hat{\bx}: (\hat{\bx},\by) \in \cS_n(\tilde{\ell}_n^{(1)}(\bx), \tilde{\ell}_n^{(2)}(\by) ) \right\} \right|}{2^{\tilde{\ell}_n^{(1)}(\bx)}} \nonumber\\
&+ \sum_{\bx,\by} P_{X^nY^n}(\bx,\by) 
 \frac{\left| \left\{ \hat{\by}: (\bx,\hat{\by}) \in \cS_n(\tilde{\ell}_n^{(1)}(\bx), \tilde{\ell}_n^{(2)}(\by) ) \right\} \right|}{2^{\tilde{\ell}_n^{(2)}(\by)}} \nonumber\\
&+ \sum_{\bx,\by} P_{X^nY^n}(\bx,\by) 
 \frac{\left| \left\{ \hat{\bx}: (\hat{\bx},\hat{\by}) \in \cS_n(\tilde{\ell}_n^{(1)}(\bx), \tilde{\ell}_n^{(2)}(\by) ) \right\} \right|}{2^{\tilde{\ell}_n^{(1)}(\bx) + \tilde{\ell}_n^{(2)}(\by)}} \\
&\le \Pr\Bigg\{ \frac{1}{n} \log \frac{1}{P_{X^n|Y^n}(X^n|Y^n)} > \frac{\ell_n^{(1)}(X^n)}{n} + \delta \text{ or } \nonumber\\
&~~~\frac{1}{n} \log \frac{1}{P_{Y^n|X^n}(Y^n|X^n)} > \frac{\ell_n^{(2)}(Y^n)}{n} + \delta \text{ or } \nonumber\\
&~~~\frac{1}{n} \log \frac{1}{P_{X^nY^n}(X^n,Y^n)} > \frac{\ell_n^{(1)}(X^n) + \ell_n^{(2)}(Y^n)}{n} + \delta \Bigg\}
 + 3 \cdot 2^{- n\delta} \\
&= P_{X^nY^n}(\cT_{n,1}^c \cup \cT_{n,2}^c \cup \cT_{n,0}^c) + 3 \cdot 2^{-n\delta}.
\end{align}
Hence, by Lemma \ref{lemma:non-typical}, we have $\mathbb{E}[\error(\hat\code_n|f_n^\id)] \to 0$ as $n\to\infty$. 
This implies that there exists a SW code $\hat\code_n = (\hvarphi_n^{(1)},\hvarphi_n^{(2)},\hpsi_n)$
satisfying \eqref{eq:theorem1-1}. 

On the other hand, the codeword length of the encoder $\hvarphi_n^{(1)}$ of the constructed SW code satisfies that 
\begin{align}
\abs{\hvarphi_n^{(1)}(\bx)} 
&\le \tilde{\ell}_n^{(1)}(\bx) + 2 ( \lfloor \log \tilde{\ell}_n^{(1)}(\bx) \rfloor + 1) \\
&\le \ell_n^{(1)}(\bx) + 2 n \delta + 2 \log (\ell_n^{(1)}(\bx) + 2 n \delta) + 3 \\
&\le \ell_n^{(1)}(\bx) + 2 n \delta + 2 \log (nL_1 + 2 n \delta) + 3.
\end{align}
Since we can take $\delta > 0$ arbitrarily small, $\hvarphi_n^{(1)}$ satisfies \eqref{eq:theorem1-2}.
Similarly, we can prove \eqref{eq:theorem1-3}. \qed

%%%%%%%%%%%%%%%%%%%%%%%%%%%%%%%%%%%%%%%%%%%%%%%%%%%%%%%%%%%
\subsection{Proof of Theorem \ref{thm:two_enc_symbolwise_fix} and Theorem \ref{thm:two_enc_symbolwise_var}}

Since ``if" part is obvious from Theorem \ref{thm:two_enc_general}, we only prove ``only if" part.
When a function is symbol-wise but not sensitive conditioned on $\cX$ or $\cY$, then
it does not satisfy Condition \ref{HK-condition1} or Condition \ref{HK-condition2} in Definition \ref{definition:HK-functions}.
Thus, the result in \cite{HanKobayashi87} implies that the Slepian-Wolf region can be improved.
Hence, it suffice to consider a symbol-wise function that is sensitive condition on $\cX$ and $\cY$, but not jointly sensitive. 

Let us consider a class of finite-state sources such that 
\begin{align}
P_{X^{2n}Y^{2n}}(\bx,\by) = \prod_{i=1}^n P_{X^2Y^2}(x_{2i-1}x_{2i}, y_{2i-1}y_{2i}).
\end{align}
In other words, let us consider a class of {\em two-symbol-wise i.i.d.} sources. Note that such a 
source $(\bX,\bY)$ includes an i.i.d. source $(\bm{U},\bm{V})$ with alphabets $\cU = \cX^2$
and $\cV = \cY^2$. 

Assume that $\bmf$ is symbol-wise but not jointly sensitive. Then, as shown in the proof of Proposition \ref{prop:HK-to-be-joint},
there exists $x^2 = (a_0,a_1)$, $\hat{x}^2 = (a_0,a_2)$, $y^2 = (b_1,b_0)$, and $\hat{y}^2 = (b_2,b_0)$ such that 
$x^2 \neq \hat{x}^2$, $y^2 \neq \hat{y}^2$, and $f_2(x^2,y^2) = f_2(\hat{x}^2,\hat{y}^2)$. Note that $f$ induces a 
function $g$ on $\cU \times \cV$ which is {\em not} an HK function.

Now, we can prove the theorem by applying the result of Han and Kobayashi \cite[Theorem 1]{HanKobayashi87}
to $(\bm{U},\bm{V})$ and $g_n(\bm{u},\bm{v}) = (g(u_1,v_1),\ldots,g(u_n,v_n))$; it should be noted that, while Han
and Kobayashi deal with fixed length coding, the SW region for fixed length coding is identical with that of 
variable length coding if the source is i.i.d.. Further, it is not hard to see that $(\bX,\bY)$ is 
smooth if $(\bm{U},\bm{V})$ satisfies the positivity condition, i.e., $P_{UV}(u,v) > 0$ for all $(u,v)$. \qed

%%%%%%%%%%%%%%%%%%%%%%%%%%%%%%%%%%%%%%%%%%%%%%%%%%%%%%%%%%%
\subsection{Proofs of Theorem \ref{thm:fullside_fix} and Theorem \ref{thm:fullside_var}}

The proof of these theorems are almost the same as that of Theorem \ref{thm:two_enc_func_to_SW}.
Thus, we only show the outline.\footnote{Again, the result for fixed-length (Theorem \ref{thm:fullside_fix}) can be proved as 
a special case of the variable-length code.} 
For a given variable-length code $\code_n = (\varphi_n^{(1)},\psi_n)$ for computing $f_n$, 
by a similar argument as Section \ref{subsection:proof-thm:two_enc_func_to_SW}, we can show that there exists 
a SW code (with full side-information) $\hat\code_n = (\hvarphi_n^{(1)}, \hpsi_n)$ satisfying 
\begin{align}
\error(\hat\code_n|f_n^\id)
&\le P_{X^nY^n}(\cT_{n,1}^c) + 2^{-n\delta} \\
&= P_{X^nY^n}(\cD_n^c \cap \cT_{n,1}^c) + P_{X^nY^n}(\cD_n \cap \cT_{n,1}^c) + 2^{-n\delta} \\
&\le \error(\code_n|f_n) + P_{X^nY^n}(\cD_n \cap \cT_{n,1}^c) + 2^{-n\delta}
\label{eq:proof-full-side-1}
\end{align} 
and 
\begin{align} \label{eq:length-of-constructed-SW-code}
\abs{\hvarphi_n^{(1)}(\bx)} \le \ell_n^{(1)}(\bx) + 2 n \delta + 2 \log (nL_1 + 2 n \delta) + 3.
\end{align}
Now, we apply Lemma \ref{lemma:core} to \eqref{eq:proof-full-side-1}, and obtain
\begin{align} \label{eq:error-probability-of-constructed-SW-code}
\error(\hat\code_n|f_n^\id) 
 \le \left(1+ \frac{2 |\cY|}{\beta q} \right) \error(\code_n|f_n) + (v_n(\beta)+2) 2^{-n\delta}.
\end{align}
Thus, by taking $\beta > 0$ sufficiently small compared to $\delta >0 $, we can derive the statement of the theorem. \qed

%%%%%%%%%%%%%%%%%%%%%%%%%%%%%%%%%%%%%%%%%%%%%%%%%%%%%%%%%%%
\subsection{Proof of Theorem \ref{thm:weakly_slowly_varying}}

The only modifications we need is the proof of Lemma \ref{lemma:core}.
In the proof of Lemma \ref{lemma:core}, we use the properties of sensitivity and smooth in
\eqref{eq:consequence-of-sensitivity} and \eqref{eq:consequence-of-slowly-varying}.
Suppose that $P_{X^nY^n}(\bx_k^\prime,\by) \cdot P_{X^nY^n}(\bx_k^{\prime\prime},\by) > 0$
(otherwise, since $P_{X^nY^n}(\bx_{v+2k},\by) = 0$, the desired inequality 
$P_{X^nY^n}(\bx_{k,j}^*,\by_j) \ge q P_{X^nY^n}(\bx_{v+2k},\by)$ holds trivially) and $\bx_k^\prime$ and $\bx_k^{\prime\prime}$
differ in $i_1$th, \ldots, $i_{\lceil \beta n \rceil}$th positions. Since $(\bX,\bY)$ is weakly smooth, 
for each $j =1,\ldots,\lceil \beta n \rceil$,
there exists $\by_j$ that differs from $\by$ only in $i_j$th position and 
\begin{align}
P_{X^nY^n}(\bx_k^\prime, \by_j) &\ge q P_{X^nY^n}(\bx_k^\prime,\by), \\
P_{X^nY^n}(\bx_k^{\prime\prime},\by_j) &\ge q P_{X^nY^n}(\bx_k^{\prime\prime},\by).
\end{align} 
Furthermore, since $f_n$ is highly sensitive conditioned on $\cY^n$, we have
$f_n(\bx_k^\prime,\by_j) \neq f_n(\bx_k^{\prime\prime},\by_j)$, which implies either of the
events in \eqref{eq:consequence-of-sensitivity}
is true. Then, by defining $\bx_{k,j}^*$ in the same manner as the proof of Lemma \ref{lemma:core}, 
\eqref{eq:consequence-of-slowly-varying} also holds.
The rest of the proof goes through exactly in the same manner. \qed

%%%%%%%%%%%%%%%%%%%%%%%%%%%%%%%%%%%%%%%%%%%%%%%%%%%%%%%%%%%
\subsection{Proof of Theorem \ref{thm:not_totally_sensitive_bounds}}\label{subsec:proof_thm_not_totally_sensitive_bounds}

The key of the proof is to modify \eqref{eq:core3} in Lemma
\ref{lemma:core} as follows: Under the assumption of the theorem, we have
\begin{align}
P_{X^nY^n}(\cD_n \cap \tilde{\cT}_{n,0}^c) &\le \frac{2(|\cX|+|\cY|) }{\beta q} \error(\code_n|f_n) + 
 [(v_n(\beta) +1) + (u_n(\beta)+1)] 2^{- n \delta} \label{eq:core3modified}
\end{align}
where
\begin{align}
 \tilde{\cT}_{n,0} &\eqtri \left\{ (\bx,\by) : \frac{1}{n} \log \frac{1}{P_{X^nY^n}(\bx,\by)} 
  \le \frac{\ell_n^{(1)}(\bx)}{n} + \frac{\ell_n^{(2)}(\by)}{n} + r + 2\delta \right\}.
\end{align}
Then, by using the same construction as the proof of Theorem
\ref{thm:two_enc_func_to_SW}, we can show that, for any $\lambda\in[0,1]$, there exists a SW code
$\hat\code_n = (\hvarphi_n^{(1)},\hvarphi_n^{(2)},\hpsi_n)$ satisfying
\eqref{eq:theorem1-1} and
\begin{align}
\abs{\hvarphi_n^{(1)}(\bx)} &\le \ell_n^{(1)}(\bx) + 2 n \delta + 2 \log
 (nL_1 + 2 n \delta) + 2 + n\lambda r,\\
\abs{\hvarphi_n^{(2)}(\by)} &\le \ell_n^{(2)}(\by) + 2 n \delta + 2 \log
 (nL_1 + 2 n \delta) + 2 + n(1-\lambda)r.
\end{align}
Hence, we have
\begin{align}
\cR^{\mathsf{**}}(\bX,\bY|\bmf)&\subseteq 
\left\{
(R_1,R_2)
:0\leq\exists\lambda\leq 1,(R_1+\lambda r,R_2+(1-\lambda)r)\in\cR_{\SW}^{\mathsf{**}}(\bX,\bY)\right\}.
\label{eq1:proof_thm_not_totally_sensitive_bounds}
\end{align}

On the other hand, note that $f_n$ is
sensitive conditioned on both of $\cX^n$ and $\cY^n$ by the definition
of $r$-total sensitivity.\footnote{
It should be noted that sensitivity conditioned on $\cX^n$ and $\cY^n$ is  not only used to derive \eqref{eq2:proof_thm_not_totally_sensitive_bounds}, but it is also used to derive \eqref{eq1:proof_thm_not_totally_sensitive_bounds} (cf.~the proof of \eqref{eq:core3modified}).  
} Hence, from Theorems \ref{thm:fullside_fix} and
\ref{thm:fullside_var}, we have
\begin{align}
\cR^{\mathsf{**}}(\bX,\bY|\bmf)&\subseteq 
\left\{
(R_1,R_2): R_1\geq R_{\SW}^{\mathsf{**}}(\bX|\bY), R_2\geq R_{\SW}^{\mathsf{**}}(\bY|\bX)
\right\}.
\label{eq2:proof_thm_not_totally_sensitive_bounds}
\end{align}

Combining \eqref{eq1:proof_thm_not_totally_sensitive_bounds} and
\eqref{eq2:proof_thm_not_totally_sensitive_bounds}, we have the theorem.

Now, we prove \eqref{eq:core3modified}. Since \eqref{eq:core3modified}
is a modification of \eqref{eq:core3}, we explain how the proof of
\eqref{eq:core3}, which is given in Appendix \ref{appendix:proof_lemma},
is modified to prove \eqref{eq:core3modified}.

Since \eqref{eq:r-totally-sensitive} holds, for sufficiently
large $n$ and for any $z_n\in\cZ_n$, we have
\begin{align}
\EQ(z_n|f_n)\leq 2^{n(r+\delta)}.
\end{align}
This guarantees that, instead of \eqref{eq:proof-core-cD_ab}, we can find
$(\bx_1,\by_1),\dots,(\bx_J,\by_J)$ such that $\bx_i\neq\bx_j$ and
$\by_i\neq\by_j$ for every $i\neq j$, $J\leq 2^{n(r+\delta)}$, and
\begin{align}
 \cD_{a,b}\subseteq\bigcup_{i=1}^J\left[ (\cD_{a,\by_i} \times \{ \by_i \}) \cup (\{ \bx_i \} \times \cD_{\bx_i,b}) \right].
\end{align}
Thus, instead of \eqref{eq:proof-core-3}, we have
\begin{align}
&P_{X^nY^n}(\cD \cap \tilde{\cT}_0^c)\nonumber\\
&\leq \sum_{a \in \cC^{(1)}} \sum_{b \in \cC^{(2)}} 
\left[
\sum_{i=1}^J\sum_{\bx\in\cD_{a,\by_i}}P_{X^nY^n}(\bx,\by_i) \bm{1}[(\bx,\by_i) \in \tilde{\cT}_0^c] 
+
\sum_{i=1}^J\sum_{\by\in\cD_{\bx_i,b}}P_{X^nY^n}(\bx_i,\by) \bm{1}[(\bx_i,\by) \in \tilde{\cT}_0^c] 
\right]\\
&\le \sum_{a \in \cC^{(1)}} \sum_{b \in \cC^{(2)}} J[ (v+1)+(u+1)] 2^{- \ell(a) - \ell(b) -r - 2n \delta}  \nonumber\\
& + \sum_{a \in \cC^{(1)}} \sum_{b \in \cC^{(2)}} \sum_{i=1}^J \sum_{k=1}^{\left[ \frac{1}{2} (|\cD_{a,\by_i}| - v )\right]^+}
  \left( P_{X^nY^n}(\bx_{v+2k},\by_i) + P_{X^nY^n}(\bx_{v+2k+1},\by_i) \right) \nonumber\\
& +   \sum_{a \in \cC^{(1)}} \sum_{b \in \cC^{(2)}} \sum_{i=1}^J \sum_{k=1}^{\left[ \frac{1}{2} (|\cD_{\bx_i,b}| - u )\right]^+}
  \left( P_{X^nY^n}(\bx_i,\by_{u+2k}) + P_{X^nY^n}(\bx_i,\by_{u+2k+1}) \right) \label{eq:proof-core-3modified}.
\end{align}
Then, each term in \eqref{eq:proof-core-3modified}
is upper bounded in the same way as 
\eqref{eq:proof-core-4}, \eqref{eq:proof-core-5}, and
\eqref{eq:proof-core-6} respectively.
Hence, we have \eqref{eq:core3modified}.
\qed

%%%%%%%%%%%%%%%%%%%%%%%%%%%%%%%%%%%%%%%%%%%%%%%%%%%%%%%%%%%
\subsection{Proof of Theorem \ref{thm:example_outer_bound}}\label{subsec:proof_thm_example_outer_bound}
Let $\mathsf{p}$ be the smallest prime integer larger than $\abs{\cX}+\abs{\cY}-2$
and consider a Galois field $\GF(\mathsf{p})=\{0,1,\dots,\mathsf{p}-1\}$.
Without loss of generality, we assume that
$\cX=\{0,1,\dots,\abs{\cX}-1\}\subseteq \GF(\mathsf{p})$ and
$\cY=\{0,1,\dots,\abs{\cY}-1\}\subseteq \GF(\mathsf{p})$.
Then, let us define the function $f_n$ as
\begin{align}
 f_n(\bx,\by)&\eqtri \left(x_1\oplus y_1, x_2\oplus y_2, \dots
 x_{n\rho}\oplus y_{n\rho}, (x_{n\rho+1},y_{n\rho+1}), (x_{n\rho+2},y_{n\rho+2}),\dots, (x_n,y_n)
\right)
\label{eq1:proof_thm_example_outer_bound}
\end{align}
where $\oplus$ is addition in $\GF(\mathsf{p})$.\footnote{ More precisely, $n\rho$ in 
\eqref{eq1:proof_thm_example_outer_bound} should be $\lfloor
n\rho\rfloor$, but we omit the floor function for the simplicity.}  In other words, the first $n\rho$
symbols of $f_n(\bx,\by)$ is symbol-wise addition in $\GF(\mathsf{p})$ and the
remaining part of $f_n(\bx,\by)$ is identical with the last $n(1-\rho)$ symbols of
$(\bx,\by)$.
We can see that $\bmf=\{f_n\}_{n=1}^\infty$ is $r$-totally sensitive,
since $\max_{z_n\in\cZ_n}\EQ(z_n|f_n)=M^{n\rho}=2^{nr}$, where $M\eqtri\min\{\abs{\cX},\abs{\cY}\}=2^{\barr}$.

On the other hand, we consider a general source $(\bX,\bY)$ defined as
follows.
Fix $\varepsilon>0$ specified later, and let $Q_{XY}$ be a joint distribution on $\cX\times\cY$ such that
\begin{align}
 Q_{XY}(x,y)&=
\begin{cases}
 \frac{1-\varepsilon}{M}& (x,y)=(0,M-1),(1,M-2),\dots,(M-1,0),\\
 \frac{\varepsilon}{\abs{\cX}\abs{\cY}-M}& \text{otherwise}.
\end{cases}
\end{align}
Then, let us define the joint distribution of $(X^n,Y^n)$ as
\begin{align}
 P_{X^nY^n}(\bx,\by)&=
\left[
\prod_{i=1}^{n\rho}Q_{XY}(x_i,y_i)\right]\times
\left(
\frac{1}{\abs{\cX}\abs{\cY}}\right)^{n(1-\rho)}
\label{eq2:proof_thm_example_outer_bound}
\end{align}
for all $(\bx,\by)$.
That is, the first $n\rho$ symbols of $(X^n,Y^n)$ is i.i.d.~with the
joint distribution $Q_{XY}$ and the last $n(1-\rho)$ symbols of
$(X^n,Y^n)$ is i.i.d.~with the uniform distribution on $\cX\times\cY$.
We can see that $(\bX,\bY)\eqtri\{(X^n,Y^n)\}_{n=1}^\infty$ is smooth, since $Q_{XY}$ satisfies the positivity condition.

Now, we prove that
\eqref{eq1:thm_example_outer_bound} and
\eqref{eq2:thm_example_outer_bound} hold  for $\bmf$ and $(\bX,\bY)$ defined above.

At first, let us construct a coding scheme for computing $f_n$ as follows:
(i) The first $n\rho$ symbols are coded by the coding scheme given in Lemma 5
of \cite{HanKobayashi87}; i.e., a generalization of the coding scheme of K\"orner and Marton
\cite{KornerMarton79}.
(ii) The remaining $n(1-\rho)$ symbols are sent to the decoder without compression.
Note that if $(X,Y)\sim Q_{XY}$ and $\varepsilon$ is
sufficiently small then
\begin{align}
H(X\oplus Y)\leq h(\varepsilon)+
\varepsilon\log\left(\abs{\cX}\abs{\cY}-M\right)\leq \delta/\rho.
\end{align}
Thus, by the coding scheme described above, 
the pair $(R_1,R_2)$
satisfying $R_1\geq\delta+(1-\rho)\log\abs{\cX}$
and $R_2\geq\delta+(1-\rho)\log\abs{\cY}$ is achievable.
Hence, we have \eqref{eq1:thm_example_outer_bound}.

On the other hand, note that if $(X,Y)\sim Q_{XY}$ and $\varepsilon$ is
sufficiently small then
\begin{align}
 H(X,Y)&=(1-\varepsilon)\log
 M+h(\varepsilon)+\varepsilon\log\left(\abs{\cX}\abs{\cY}-M\right)\\
&\geq \barr-\delta/\rho.
\end{align}
Hence, from \eqref{eq2:proof_thm_example_outer_bound}, we have
\begin{align}
 H(X^n|Y^n)&\geq n(1-\rho)\log\abs{\cX}\label{eq3:proof_thm_example_outer_bound}\\
 H(Y^n|X^n)&\geq n(1-\rho)\log\abs{\cX}
\end{align}
and
\begin{align}
  H(X^n,Y^n)&\geq n\rho (\barr-\delta/\rho)+n(1-\rho)\log\abs{\cX}\abs{\cY}\\
&= n\left\{r-\delta+(1-\rho)\log\abs{\cX}\abs{\cY}\right\}.
\label{eq4:proof_thm_example_outer_bound}
\end{align}
From
\eqref{eq3:proof_thm_example_outer_bound}--\eqref{eq4:proof_thm_example_outer_bound},
it is not hard to see that \eqref{eq2:thm_example_outer_bound} holds.
\qed

%%%%%%%%%%%%%%%%%%%%%%%%%%%%%%%%%%%%%%%%%%%%%%%%%%%%%%%%%%%
\subsection{Proof of Theorem \ref{thm_moderate_deviation}}\label{subsec:proof_thm_moderate_deviation}

We only prove \eqref{eq:moderate-vl} since \eqref{eq:moderate-fl} can be proved in a similar manner.
It suffice to prove only one direction, i.e., $e^\vl(t,\gamma|\bmf) \le e_{SW}^\vl(t,\gamma)$.
For a given code $\{ \code_n \}_{n=1}^\infty$ satisfying \eqref{eq:moderate-rate-constraint} and
\begin{align}
\liminf_{n\to\infty} -\frac{1}{n^{1-2t}}\log \error(\code_n|f_n) \ge e^\vl(t,\gamma|\bmf),
\end{align}
we can construct a SW code $\hat\code_n = (\hvarphi_n^{(1)}, \hpsi_n)$ satisfying \eqref{eq:length-of-constructed-SW-code} 
and \eqref{eq:error-probability-of-constructed-SW-code}. We set $\beta = \beta_n = \frac{1}{n}$
and $\delta = \delta_n = \frac{1}{n^{3t/2}}$.
Then, by noting $h(\beta) \le 2\beta + 2\beta \log (1/2\beta)$ for $0 < \beta < 1/2$, we have
$v_n(\beta_n) \le 16 |\cX| n^3$. Thus, we have
\begin{align}
\limsup_{n\to\infty} \frac{1}{n^{1-t}} \left( \Ex\left[\abs{
\hvarphi_n^{(1)}(X^n)
}\right] - n H(X|Y) \right) \leq \gamma
\end{align}
and 
\begin{align}
\liminf_{n\to\infty} -\frac{1}{n^{1-2t}}\log \error(\hat\code_n|f_n^{\id}) \ge e^\vl(t,\gamma|\bmf).
\end{align}

%% file: conclusion.tex
\section{Conclusion}\label{sec:conclusion}
In this paper, we investigated a dichotomy of functions in distributed
coding: for a sequence $\bmf$ of functions, does the achievable rate
region for computing $\bmf$ coincide with the SW region?  We introduced
the class of smooth sources and gave a sufficient condition for
the coincidence: if $\bmf$ is totally sensitive then the achievable rate
region for computing $\bmf$ coincides with the SW region for any smooth sources.  Further, we proved that, for symbol-wise functions,
the total sensitivity is the necessary and sufficient condition for the
coincidence of two regions.  On the other hand, it remains as a future
work to establish the necessary and sufficient condition on functions
which may not be symbol-wise.

Moreover, as a generalization of our dichotomy theorem, we give an outer
bound on the achievable rate region for computing a class of functions
wider than the totally sensitive functions.  Of course, to characterize
the achievable rate region for general functions remains as a future
work.

In our investigation, we used the information-spectrum approach so that
we can establish the results in a unified way.  This approach allows us
to derive a refined result in the moderate deviation regime as given in
Section \ref{subsec:theorems_moderate_deviation}.  Although we consider
only i.i.d.~sources in Section \ref{subsec:theorems_moderate_deviation}
for simplicity, it is not hard to generalize Theorem
\ref{thm_moderate_deviation} for wider classes of sources. Indeed, the
assumption of i.i.d.~is not so critical in the proof of Theorem
\ref{thm_moderate_deviation} given in Section
\ref{subsec:proof_thm_moderate_deviation}.  On the other hand, for general
sources that have memory and may not be stationary nor ergodic, to
characterize $e_{\SW}^\vl(t,\gamma)$ and $e_{\SW}^\fl(t,\gamma)$ itself
remains as an important work.

In this paper, we considered only lossless computation, where the error
probability is required to tend zero as the block size goes to infinity.
It is an important future work to generalize our results for
$\varepsilon$-error case, where the error probability is required only
to be smaller than the given threshold $\varepsilon>0$.  When we
consider $\varepsilon$-error case, the strong converse property is an
important subject to be investigated; e.g., it is an interesting problem
to establish the necessary and sufficient condition on functions so that
the strong converse holds for function computation whenever the strong
converse holds for SW coding.  Furthermore, it is also an important
future work to generalize our results for lossy case and to establish
the condition so that the rate-distortion region for distributed
computing coincides with that for distributed source coding.

%% file: appendix_proof_lemma.tex
\section{Proof of Proposition \ref{prop:HK-to-be-joint}}\label{appendix:proof_prop_HK-to-be-joint}
\paragraph{If part}

At first, we assume that $f$ is an HK function and satisfies 1) of the proposition.
Then we have 
\begin{equation}
 f(a_1,b_1)=f(a_2,b_2) \text{ means } b_1=b_2.\label{eq:proof_prop_HK-to-be-join}
\end{equation}
Indeed, if $a_1= a_2$ then \eqref{eq:proof_prop_HK-to-be-join} follows from 1) of the proposition. 
Moreover, if $a_1\neq a_2$ then \eqref{eq:proof_prop_HK-to-be-join} follows from the condition
3) in the definition of HK functions.

Now, note that if $f_n(\bx,\by)=f_n(\hbx,\hby)$
then $f(x_i,y_i)=f(\hx_i,\hy_i)$ for all $i=1,2,\dots, n$, 
since $f_n$ is symbol-wise.
Hence, by \eqref{eq:proof_prop_HK-to-be-join}, we can see that 
if $f_n(\bx,\by)=f_n(\hbx,\hby)$ then $y_i=\hy_i$ for all $i=1,2,\dots,n$, that is, $\by=\hby$.

On the other hand, similar argument holds for a case where $f$ satisfies 2) of the
proposition, and we can show that if $f_n(\bx,\by)=f_n(\hbx,\hby)$ then
$\bx=\hbx$ in this case.

Summarizing the above, 
if $f$ is an HK function and satisfies 1) or 2) of the proposition then
$f_n(\bx,\by)=f_n(\hbx,\hby)$ implies $\bx=\hbx$ or $\by=\hby$.
This completes the proof of ``if part''.
\qed

\paragraph{Only if part}
We prove this part by contradiction. If
$f$ does not satisfies 1)  then there exists
$b_1,b_2\in\cY$ and $a_0\in\cX$ such that $b_1\neq b_2$ and
$f(a_0,b_1)=f(a_0,b_2)$. 
Similarly, if $f$ does not satisfies 2)  then there exists $a_1,a_2\in\cX$ and
$b_0\in\cY$ such that $a_1\neq a_2$ and $f(a_1,b_0)=f(a_2,b_0)$. Hence, if $f$ does not satisfies 1) nor 2)
then $x^2=(a_0,a_1)$, $\hx^2=(a_0,a_2)$, $y^2=(b_1,b_0)$,
and $\hy^2=(b_2,b_0)$ satisfy $x^2\neq \hx^2$, $y^2\neq \hy^2$, and
$f_2(x^2,y^2)=f_2(\hx^2,\hy^2)$.
\qed

\section{Proof of Proposition \ref{prop:AC_weak_condition}}\label{appendix:proof_AC_weak_condition}

\paragraph{If part}
Let $q\eqtri \min\{P_{XY}(a,b): (a,b)\in\cX\times\cY, P_{XY}(a,b)>0\}$.
Fix
$\bx\neq\hbx$ and $\by$ satisfying $P_{X^nY^n}(\bx,\by)\cdot P_{X^nY^n}(\hbx,\by)>0$
arbitrarily, and suppose that $x_i\neq\hx_i$. 
Since $P_{X_iY_i}(x_i,y_i)\cdot P_{X_iY_i}(\hx_i,y_i)>0$ holds, by the
assumption, there exists
$b\neq y_i$ satisfying $P_{X_iY_i}(x_i,b)\cdot P_{X_iY_i}(\hx_i,b)>0$.
We can see that 
$\hby\in\cY^n$ obtained by replacing the $i$th component of $\by$ with
$b$ satisfies
\eqref{eq1:def_weakly_slowly_varying} and \eqref{eq2:def_weakly_slowly_varying}.
\qed

\paragraph{Only if part}
This part is obvious, since 
if the source is weakly smooth then
the property required in 
Definition \ref{def:weakly-slowly-varying}
holds for $n=1$.
\qed

%%%%%%%%%%%%%%%%%%%%%%%%%%%%%%%%%%%%%%%%%%%%%%%%%%%%
\section{Proof of Lemma \ref{lemma:core}}\label{appendix:proof_lemma}

Throughout the proof, we omit subscript $n$ if it is obvious from the context. Furthermore, we also omit $\beta$ from
$v_n(\beta)$ and $u_n(\beta)$, and thus they are just denoted by $v$ and $u$. 
For $a \in \cC^{(1)}$ and $b \in \cC^{(2)}$, let
\begin{align}
\cD_{a,b} &\eqtri \left\{ (\bx,\by) : \varphi^{(1)}(\bx) = a, \varphi^{(2)}(\by) = b, (\bx,\by) \in \cD \right\}, \\
\cD_{a,\by} &\eqtri \left\{ \bx : \varphi^{(1)}(\bx) = a, (\bx,\by) \in \cD \right\}, \\
\cD_{\bx,b} &\eqtri \left\{ \by : \varphi^{(2)}(\by) = b, (\bx,\by) \in \cD \right\}.
\end{align}

\paragraph*{Proof of \eqref{eq:core1}}
We leverage El Gamal's argument \cite{ElGamal83}.
For each $(a,\by)$, we sort the elements in $\cD_{a,\by}$ in the decreasing order of probabilities, i.e.,
\begin{align} \label{eq:sorting}
P_{X^n Y^n}(\bx_1,\by) \ge P_{X^nY^n}(\bx_2,\by) \ge \cdots \ge P_{X^nY^n}(\bx_{|\cD_{a,\by}|},\by).
\end{align}
First, we take $\bx_1^\prime \eqtri \bx_1$, and pair it with an $\bx_1^{\prime\prime} \in \cD_{a,\by}$ that satisfies
$d(\bx_1^\prime, \bx_1^{\prime\prime}) \ge \beta n$ and has the largest probability. 
Clearly, we have
\begin{align}
P_{X^n Y^n}(\bx_1^{\prime\prime},\by) \ge P_{X^nY^n}(\bx_{v+2},\by).
\end{align}
Next, we select the $\bx_2^\prime \in \cD_{a,\by} \backslash \{ \bx_1^\prime,\bx_1^{\prime\prime} \}$ with
the largest probability, and pair it with an unselected $\bx_2^{\prime\prime}$ satisfying 
$d(\bx_2^\prime,\bx_2^{\prime\prime}) \ge \beta n$ and that has the largest probability. Clearly, we have
\begin{align}
P_{X^nY^n}(\bx_2^\prime,\by) &\ge P_{X^nY^n}(\bx_3,\by), \label{eq:bound-pairing-1} \\
P_{X^nY^n}(\bx_2^{\prime\prime},\by) &\ge P_{X^nY^n}(\bx_{v+4},\by). \label{eq:bound-pairing-2}
\end{align}
We repeat this process until no more pairing is possible.\footnote{This process continues at least 
$\left[ \frac{1}{2} (|\cD_{a,\by}| - v )\right]^+$ times, which may be $0$.}
Then, since $f_n$ is sensitive conditioned on $\cY^n$, for each pair $(\bx_k^\prime,\bx_k^{\prime\prime})$, we can find
$\by_{k,1},\ldots,\by_{k,\lceil \beta n\rceil}$ such that $d(\by, \by_{k,j}) = 1$ and 
$f_n(\bx_k^\prime,\by_{k,j}) \neq f_n(\bx_k^{\prime\prime},\by_{k,j})$, which implies that either 
\begin{align} \label{eq:consequence-of-sensitivity}
(\bx_k^\prime,\by_{k,j}) \in \cD^c \mbox{ or } (\bx_k^{\prime\prime}, \by_{k,j}) \in \cD^c
\end{align}
is true. For each $j$, let $\bx_{k,j}^* \in \{ \bx_k^{\prime}, \bx_k^{\prime\prime} \}$ be such that $(\bx_{k,j}^*,\by_{k,j}) \in \cD^c$.
Since $(\bX,\bY)$ is smooth with respect to $\bY$, we have
\begin{align}
P_{X^nY^n}(\bx_{k,j}^*,\by_{k,j}) &\ge q P_{X^nY^n}(\bx_{k,j}^*,\by) \label{eq:consequence-of-slowly-varying} \\
&\ge q P_{X^nY^n}(\bx_{v+2k},\by),
\end{align} 
where the second inequality follows from the procedure of pairing (cf.~\eqref{eq:bound-pairing-1} and \eqref{eq:bound-pairing-2}).
Thus, we have
\begin{align}
\lceil \beta n \rceil q \sum_{k=1}^{\left[ \frac{1}{2} (|\cD_{a,\by}| - v )\right]^+} P_{X^nY^n}(\bx_{v+2k},\by) 
 \le \sum_{j=1}^{\lceil \beta n \rceil} \sum_{k=1}^{\left[ \frac{1}{2} (|\cD_{a,\by}| - v )\right]^+} P_{X^nY^n}(\bx_{k,j}^*,\by_{k,j}).
\end{align}
Here,\footnote{It should be noted that $\bx_{k,j}^*$ and $\by_{k,j}$ implicitly depend on $a$ and $\by$.} note that 
\begin{align}
\bigcup_{a \in \cC^{(1)}} \bigcup_{\by \in \cY^n} \left\{ (\bx_{k,j}^*, \by_{k,j}) : k = 1,\ldots, \left[ \frac{1}{2} (|\cD_{a,\by}| - v )\right]^+,
 j=1,\ldots,\lceil \beta n \rceil \right\}
\subseteq \cD^c,
\end{align}
and each element in $\cD^c$ overlaps at most $n|\cY|$ times in the lefthand side. Thus, we have
\begin{align}
\sum_{a \in \cC^{(1)}} \sum_{\by} \sum_{k=1}^{\left[ \frac{1}{2} (|\cD_{a,\by}| - v )\right]^+} P_{X^nY^n}(\bx_{v+2k},\by) 
 &\le \frac{ n |\cY|}{\beta n q} \sum_{(\bx,\by) \in \cD^c} P_{X^nY^n}(\bx,\by)  \\
 &= \frac{ |\cY|}{\beta q} \error(\code_n|f_n).
 \label{eq:appendix-bound-C1}
\end{align}

Now, we have 
\begin{align}
P_{X^nY^n}(\cD \cap \cT_1^c) 
&= \sum_{a \in \cC^{(1)}} \sum_{\by} \sum_{\bx \in \cD_{a,\by}} P_{X^nY^n}(\bx,\by) \bm{1}[(\bx,\by) \in \cT_1^c] \\
&\le \sum_{a \in \cC^{(1)}} \sum_{\by} (v+1) P_{Y^n}(\by) 2^{- \ell(a) - n\delta}  \nonumber\\
& + \sum_{a \in \cC^{(1)}} \sum_{\by} \sum_{k=1}^{\left[ \frac{1}{2} (|\cD_{a,\by}| - v )\right]^+} 
  \left( P_{X^nY^n}(\bx_{v+2k},\by) + P_{X^nY^n}(\bx_{v+2k+1},\by) \right) \label{eq:proof-core-1}\\
&\le (v+1) 2^{- n \delta} + 2 \sum_{a \in \cC^{(1)}} \sum_{\by} \sum_{k=1}^{\left[ \frac{1}{2} (|\cD_{a,\by}| - v )\right]^+} 
  P_{X^nY^n}(\bx_{v+2k},\by) \\
&\le (v+1) 2^{- n \delta} + \frac{2 |\cY|}{\beta q} \error(\code_n|f_n),
\end{align}
where $\ell(a)$ is the length of codeword $a$; the first inequality is derived by splitting $\cD_{a,\by}$ into
the first $(v+1)$ elements and the rest, and then by applying the property of $\cT_1^c$ to the former;
and the second inequality follows from the Kraft inequality.
Thus, we have the desired bound. The bound \eqref{eq:core2} is proved exactly in the same manner. 
\qed

%%%%%%%%%%%%%%%%%%%%%%%%%%%%%%%%%%%%%%%%%%%%%%%%%%%%
\paragraph*{Proof of \eqref{eq:core3}}

To bound $P_{X^nY^n}(\cD \cap \cT_0^c)$, we need the following observation.
Since $f_n$ is jointly sensitive, if we pick arbitrary $(\bx_{a,b}^*,\by_{a,b}^*) \in \cD_{a,b}$, the 
following must be true:
\begin{align}
\cD_{a,b} \subseteq (\cD_{a,\by_{a,b}^*} \times \{ \by_{a,b}^* \}) \cup (\{\bx_{a,b}^* \} \times \cD_{\bx_{a,b}^*, b} ).
\label{eq:proof-core-cD_ab}
\end{align}
Otherwise, there exists $(\bx,\by) \in \cD_{a,b}$ such that 
$\bx \neq \bx_{a,b}^*$ and $\by \neq \by^*_{a,b}$, but it contradict with the fact that 
$f_n$ is jointly sensitive.\footnote{In fact, joint sensitivity of $f_n$ implies a sightly stronger statement, that is,
one of the following must be true:
\begin{align*}
\cD_{a,b} = \cD_{a,\by_{a,b}^*} \times \{ \by_{a,b}^* \} \mbox{ or }
\cD_{a,b} = \{ \bx_{a,b}^* \} \times \cD_{\bx_{a,b}^*, b}.
\end{align*}} Consequently, we have
\begin{align}
P_{X^nY^n}(\cD \cap \cT_0^c)
&= \sum_{a \in \cC^{(1)}} \sum_{b \in \cC^{(2)}} \sum_{(\bx,\by) \in \cD_{a,b}} P_{X^nY^n}(\bx,\by) \bm{1}[(\bx,\by) \in \cT_0^c] \\
&\le \sum_{a \in \cC^{(1)}} \sum_{b \in \cC^{(2)}}
 \Bigg[ \sum_{\bx \in \cD_{a,\by_{a,b}^*}} P_{X^nY^n}(\bx,\by_{a,b}^*) \bm{1}[(\bx,\by_{a,b}^*) \in \cT_0^c ] \nonumber\\
&  + \sum_{\by \in \cD_{\bx_{a,b}^*,b}} P_{X^nY^n}(\bx_{a,b}^*,\by) \bm{1}[(\bx_{a,b}^*,\by) \in \cT_0^c] \Bigg] \\
&\le \sum_{a \in \cC^{(1)}} \sum_{b \in \cC^{(2)}} [ (v+1)+(u+1)] 2^{- \ell(a) - \ell(b) - n \delta}  \nonumber\\
& + \sum_{a \in \cC^{(1)}} \sum_{b \in \cC^{(2)}} \sum_{k=1}^{\left[ \frac{1}{2} (|\cD_{a,\by_{a,b}^*}| - v )\right]^+}
  \left( P_{X^nY^n}(\bx_{v+2k},\by_{a,b}^*) + P_{X^nY^n}(\bx_{v+2k+1},\by_{a,b}^*) \right) \nonumber\\
& +   \sum_{a \in \cC^{(1)}} \sum_{b \in \cC^{(2)}} \sum_{k=1}^{\left[ \frac{1}{2} (|\cD_{\bx_{a,b}^*,b}| - u )\right]^+}
  \left( P_{X^nY^n}(\bx_{a,b}^*,\by_{u+2k}) + P_{X^nY^n}(\bx_{a,b}^*,\by_{u+2k+1}) \right) \label{eq:proof-core-3}, 
\end{align}
where $\by_{u+2k}$ is defined in a similar manner as $\bx_{v+2k}$ by sorting the elements in
$\cD_{\bx,b}$ for each $\bx$ and $b$ (cf.~\eqref{eq:sorting}), and where the inequality in
\eqref{eq:proof-core-3} is derived in a similar manner as the inequality in \eqref{eq:proof-core-1}.
By the Kraft inequality, we have
\begin{align}
\sum_{a \in \cC^{(1)}} \sum_{b \in \cC^{(2)}} [ (v+1)+(u+1)] 2^{- \ell(a) - \ell(b) - n \delta} \le [ (v+1)+(u+1)] 2^{- n \delta}.\label{eq:proof-core-4}
\end{align}
By using \eqref{eq:appendix-bound-C1}, we have
\begin{align}
& \sum_{a \in \cC^{(1)}} \sum_{b \in \cC^{(2)}} \sum_{k=1}^{\left[ \frac{1}{2} (|\cD_{a,\by_{a,b}^*}| - v )\right]^+}
  \left( P_{X^nY^n}(\bx_{v+2k},\by_{a,b}^*) + P_{X^nY^n}(\bx_{v+2k+1},\by_{a,b}^*) \right) \nonumber\\
&\le 2 \sum_{a \in \cC^{(1)}} \sum_{\by} \sum_{k=1}^{\left[ \frac{1}{2} (|\cD_{a,\by}| - v )\right]^+} P_{X^nY^n}(\bx_{v+2k},\by)  \\
&\le \frac{2 |\cY|}{\beta q} \error(\code_n|f_n).\label{eq:proof-core-5}
\end{align}
Similarly, we have
\begin{align}
& \sum_{a \in \cC^{(1)}} \sum_{b \in \cC^{(2)}} \sum_{k=1}^{\left[ \frac{1}{2} (|\cD_{\bx_{a,b}^*,b}| - u )\right]^+}
  \left( P_{X^nY^n}(\bx_{a,b}^*,\by_{u+2k}) + P_{X^nY^n}(\bx_{a,b}^*,\by_{u+2k+1}) \right)  \nonumber\\
&\le \frac{2 |\cX|}{\beta q} \error(\code_n|f_n).\label{eq:proof-core-6}
\end{align}
Thus, we have the desired bound. \qed